\documentclass{article}

\usepackage{amsmath}
\usepackage{array,multirow,graphicx}
\usepackage{rotating}
\usepackage{makecell}
\usepackage[colorlinks=true, allcolors=black]{hyperref}
\usepackage{booktabs}
\usepackage[a4paper,top=2cm,bottom=2cm,left=3cm,right=3cm,marginparwidth=1.75cm]{geometry}
\usepackage{pifont}
\newcommand{\cmark}{\ding{51}}%
\newcommand{\xmark}{\ding{55}}%

\usepackage{comment}

\usepackage{pgfplots}
\usepackage{pgfplotstable}
\usepackage{float}
\usepackage[numbers]{natbib}
\usepackage{authblk}
\usepackage[untagged, highstructure]{accessibility}
\usepackage{changepage}

\usepackage{enumitem,amssymb}
\newlist{checkboxes}{itemize}{2}
\setlist[checkboxes]{label=$\square$}

\begin{document}

\title{Invisible, Unreadable, and Inaudible Cookie Notices: An Evaluation of Cookie Notices for Users with Visual Impairments}
\author[1]{James M Clarke}
\author[2]{Maryam Mehrnezhad}
\author[1]{Ehsan Toreini}
\affil[1]{University of Surrey, J.M.Clarke@surrey.ac.uk \& E.Toreini@surrey.acuk}
\affil[2]{Royal Holloway University of London, Maryam.Mehrnezhad@rhul.ac.uk}
\date{}                     
\setcounter{Maxaffil}{0}
\renewcommand\Affilfont{\itshape\small}

\maketitle

\begin{abstract}
This paper investigates the accessibility of cookie notices on websites for users with visual impairments (VI) via a set of system studies on top UK websites (n=46) and a user study (n=100). We use a set of methods and tools--including accessibility testing tools, text-only browsers, and screen readers, to perform our system studies.
Our results demonstrate that the majority of cookie notices on these websites have some form of accessibility issue, including contrast issues, not having headings, and not being read aloud immediately when the page is loaded. We discuss how such practices impact the user experience and privacy and provide a set of recommendations for multiple stakeholders for more accessible websites and better privacy practices for users with VIs.
To complement our technical contribution we conduct a user study, finding that people with VIs generally have a negative view of cookie notices and believe our recommendations could help their online experience.
\end{abstract}

\section{Introduction}
Visual impairment (VI) is a term used to describe any type of vision loss, ranging from partial vision loss to someone who cannot see at all \cite{adeyemiChallengesAdaptationsPublic2022,ahmedPrivacyConcernsBehaviors2015}. People with VI have various types of assistive technologies (AT) available to help them browse the internet \cite{southwellEvaluationFindingAid2013}, e.g., text-only browsers and screen readers. 
Screen readers are installed on users' computers or phones to read information by outputting it as sound \cite{rotardTactileWebBrowsing2008,southwellEvaluationFindingAid2013,wangAccessiblePrivacy2022}. They work with the browser and interpret the code that is used to build web pages \cite{chandrashekarHearingBelievingPerception}.
Screen readers are not capable of conveying visual and spatial information, such as layout and images, to the user unless relevant meta-information is provided in the web page code through \emph{markups}. 
In addition, the content of a web page is spoken out loud in a linear order, which can differ from the visual order on the screen as well as make it harder to get an overview of the page~\cite{southwellEvaluationFindingAid2013}.
To combat this, screen readers can navigate using headings to jump to different sections and they can also search for content within a web page \cite{rotardTactileWebBrowsing2008}.
This method of presentation makes it more difficult to understand the relations of a website's different parts and identify navigation links.
To ensure that AT can correctly interpret websites, there are various accessibility standards, such as the Web Content Accessibility Guidelines (WCAG) provided by the World Wide Web Consortium (W3C)  \cite{w3WCAGOverview2022}. WCAG aims to provide a shared standard for web content accessibility.
The WCAG documents explain how to make web content more accessible to disabled people. To be included in the WCAG, issues must impact disabled people with greater effect than those without disabilities \cite{w3UnderstandingConformance}.

In addition to the challenges faced solely by individuals with VIs navigating the web using AT, online tracking poses a different set of concerns for all users.
The majority of websites employ some type of tracking, using various techniques such as cookies and fingerprinting \cite{englehardtOnlineTracking1millionsite2016}. There are two types of cookies, functional and non-functional \cite{kretschmerCookieBannersPrivacy2021}, with the most common use of non-functional cookies being for personalised advertising \cite{GDPRwhatHappened,sorensenGDPRChangesThird2019,trevisanYearsEUCookie2019}. A simple method to counteract this type of tracking is to allow users to manage which cookies are stored on their device \cite{senicarPrivacyEnhancingTechnologiesApproaches2003}. With the implementation of the General Data Protection Regulation (GDPR) in 2018, companies operating in the EU and UK and/or handling EU/UK citizens' data need to choose a legal basis to collect and process user data \cite{GeneralDataProtection}.
One of the most well-known of these is cookie notices to gain consent from users~\cite{matte2020cookie}.  
Alongside the GDPR, the ePrivacy Directive and the Information Commissioner Office (ICO) give specific guidance on obtaining consent through cookie notices \cite{Directive2002582002,mehrnezhadCrossPlatformEvaluationPrivacy2020}.

Previous research has shown that individuals want to protect themselves from online tracking \cite{coopamootooFeelInvadedAnnoyed, raoWhatTheyKnow2015}, though they are not always confident~\cite{coopamootooFeelInvadedAnnoyed,mehrnezhadHowCanWould2022}. Multiple studies have looked at how the function and presentation of cookie notices differ \cite{mehrnezhadHowCanWould2022, mehrnezhadCrossPlatformEvaluationPrivacy2020, huCharacterisingThirdParty2019, utzInformedConsentStudying2019, degelingWeValueYour2019}. Similarly, studies have shown that the designs of cookie notices can affect users' interactions \cite{utzInformedConsentStudying2019}, including through dark patterns \cite{nouwensDarkPatternsGDPR2020}. Previous research has examined the effect that the GDPR and cookie notices had on the number of cookies used by websites and found in users' browsers~\cite{kretschmerCookieBannersPrivacy2021, degelingWeValueYour2019, huCharacterisingThirdParty2019}.
It has been shown that there is a disparity between the requirement of data protection laws, the practices of websites, and users' behaviour regarding online tracking protection~\cite{mehrnezhadHowCanWould2022}.

Limited research has been conducted on privacy and VIs.
Users with VIs have previously been found to have concerns about being tracked online \cite{inanInternetUseCybersecurity2016}, similar to other users \cite{coopamootooFeelInvadedAnnoyed,raoWhatTheyKnow2015}.
There has also been research looking at VI and online information credibility \cite{chandrashekarHearingBelievingPerception}. In the context of cookie notices and VI, research is extremely sparse \cite{schnellWebsitePrivacyNotification2021}. 
There are some reports on usability issues with cookie notices while looking at the wider accessibility of websites \cite{washington2019proper}. 
To the best of our knowledge, there is no research on cookie notices and AT where a comprehensive range of methods is utilised.
Our research questions include: 
\begin{itemize}
\item 
\textbf{RQ1}: How do websites and cookie notices comply with the web content accessibility guidelines and the general data protection regulations?
RQ1-a: How do popular websites comply with the current accessibility guidelines (e.g., WCAG) and the GDPR? 
RQ1-b: Does compliance necessarily mean good privacy practices for VI users? 
\item
\textbf{RQ2}: Can the existing automated accessibility tools evaluate cookie notices?
RQ2-a: How do the current cookie notices score with the automated accessibility tools (e.g., WAVE and Google Lighthouse)?
RQ2-b: Does a high score necessarily mean good practice for VI users? 
\item 
\textbf{RQ3}: How do cookie notices interface with AT? 
RQ3-a: How does the mainstream AT (e.g., text-only browsers and screen readers) interact with cookie notices?
RQ3-b: How do the current practices impact VI users' privacy?
\item 
\textbf{RQ4}: What are the general perceptions and practices of VI users regarding cookie notices?
RQ4-a: What issues have VI users encountered with cookie notices?
RQ4-b: Who do participants believe is responsible for online accessibility?
\end{itemize}

This paper contributes to the body of knowledge via its system studies, user studies, and the discussions and recommendations that we provide for improving the online privacy of users with VIs.
First, we provide a set of evaluation methods based on the off-the-shelf tools and ATs for users with VI. This enables us and other researchers to conduct system experiments and assess websites and cookie notices for their accessibility. Second, using these methods and tools, we run experiments on 46 popular UK websites (according to Alexa) and report a wide range of accessibility issues with their cookie notices. Table \ref{tab:overall} presents an overview of our system studies.
Third, we conduct user studies with 100 UK participants who use AT and extract their perceptions, practices, and preferences regarding cookies notices on websites. 
The results of our systems studies as well as the user studies confirm that current practices are far from ideal in protecting the privacy of users with VIs. 
Finally, we discuss the impact of these practices on user privacy and provide recommendations for web developers, AT designers, policymakers, and end users to improve real-world privacy practices.   

\section{Background and Related Work}
Different disabilities, including perceptional disabilities (e.g., hearing and visual), motor disabilities (i.e., limited or no use of hands, arms, legs, or mouth), and cognitive and intellectual disabilities, impact user experiences with technology and impact users in different ways~\cite{lazar2017research}.
Differential Vulnerabilities recognise how different populations face different types and degrees of security and privacy risks~\cite{pierceDifferentialVulnerabilitiesDiversity2018}. 
This challenges the universalising tendencies that frame cybersecurity around an abstract or generic user who either does not exist or is only a subset of actual end users~\cite{egelmanMythAverageUser2015,mehrnezhadCaringIntimateData2021}. 
This ties into social sciences research looking at disability and technology.
For example, the \emph{medical model} views people with disabilities as lacking something as compared to non-disabled people \cite{davidson_work_2006}, and therefore they must be cured or rehabilitated. This leads to medicalisation and a near-exclusive focus on biomedical cures~\cite{lewis_mad_2006}. This compares to the \emph{social model} which argues that it is society which disables people~\cite{union_of_physically_impaired_against_segregation_fundamental_1975}, however, this view neglects that impairment is an important aspect of many disabled people's lives and identities ~\cite{shakespeareSocialModelDisability2006}. Whereas \emph{critical realism} takes the multiple layers which make up the disabled experience into account~\cite{frauenbergerDisabilityTechnologyCritical2015}. This approach makes room for a broad range of disabled experiences, for example, personal attitudes, chronic illness, the diversity and severity of impairments, or social disadvantages. However, it still points programmatically to different ways that the lives of people with disabilities can be improved.
Both critical realism and differential vulnerabilities consider the real-world lived experiences of disabled people, as well as their thoughts. With differential vulnerabilities considering how different threats can arise for different user groups.
Studying and evaluating the privacy of users with VIs is challenging. The common range of privacy assessment methods would not be directly useful here. Instead, such approaches should be combined with accessibility assessment methods, as defined in the accessible writing guidelines of the Association for Computing Machinery \cite{hansonWritingAccessibility2015}. 

According to the Office of National Statistics in 2020, almost 11 million adults with disabilities recently used the Internet in the UK~\cite{officefornationalstatisticsInternetUsersUK}. 
A 2016 survey conducted by GOV.UK involving 712 AT users, found that 29\% utilised screen readers for Internet browsing with the remaining respondents employing various tools, such as screen magnifiers, speech recognition, or readability software \cite{gov.ukResults2016GOV2016}. This survey also studied the different screen readers in use, finding the most popular to be JAWS. WebAIM also found that JAWS was the most popular screen reader with 53.7\% of users using it as their primary screen reader and NVDA was the second most popular with 30.7\% of users using it as their primary screen reader \cite{webaimWebAIMScreenReader2021}.

\subsection{Users with VI and Privacy}

Evaluating the accessibility of websites is possible through a number of automatic and manual methods, for example, the use of tools such as screen readers and text-only browsers. Southwell and Slater have previously used the WCAG to evaluate university library finding aids \cite{southwellEvaluationFindingAid2013}. They used an automated web-accessibility checker, WAVE 5.0 by WebAim, to perform an initial assessment of each finding aid and then manually tested each website using the WebbIE 3 Web browser, which is a non-graphical text-only browser. They also used screen readers directed by keyboard navigation including \emph{System Access to GO} and \emph{NVDA}.
When using the automated checker, they found that most of the websites tested (58 of 65) had at least one accessibility error. The most common errors were missing document language, missing alternative text, missing form labels, and linked images missing alternative text. They then used the non-graphical browser, finding only 68\% had headings that enabled navigation to another important section of the document. Of those which had enough headings, they did not always have the headings in proper sequential order or were missing first-level headings. Fewer sites offered links for navigation, 57\% did, 43\% did not, and 25\% of the sites lacked both headings and links for any kind of navigation. Using the screen readers, they found that the main content of all 65 finding aids was readable; this opposes the 89\% error rate noted by the automatic checker. 

\begin{table*}[t]
    \begin{adjustwidth}{-.5in}{-.5in}  
    \centering
    \footnotesize 
    \caption{Overall view of our system studies }
    \begin{tabular}{l|c|c|c|c}
    \hline
        \multirow{3}{*}{\textbf{Method}}
        & \textbf{(I) Cookie notice} 
        & \textbf{(II) General} 
        & \textbf{(III) Manual}  
        & \textbf{(IV) Manual} \\
        & \textbf{\& Tracking} &\textbf{Automated}
        &\textbf{Testing via}
        &\textbf{Testing via}\\
        & \textbf{Behaviour Evaluation} & \textbf{Accessibility Tools} & \textbf{Text-only Browser} &  \textbf{Screen Readers}\\\hline
        \multirow{2}{*}{Tools} & \multirow{2}{*}{Google Chrome, Brave} &WAVE, & \multirow{2}{*}{WebbIE} & \multirow{2}{*}{JAWS, NVDA}\\
        & & Google Lighthouse & & \\\hline
        Website Accessibility &\multirow{2}{*}{NA} & \multirow{2}{*}{Yes} &\multirow{2}{*}{Yes}&\multirow{2}{*}{Yes}\\
        Assessment &  & & \\\hline
        Cookie Notice  &\multirow{2}{*}{Yes (General, Baseline)}&\multirow{2}{*}{Partial (Accessibility)}&\multirow{2}{*}{Yes (Accessibility)}&\multirow{2}{*}{Yes (Accessibility)}\\
        Assessment & & & & \\\hline
    \end{tabular}
    \label{tab:overall}
    \end{adjustwidth}
\end{table*}

There is scarce research on security and privacy relating to users with VI. Brulé et al. analysed 178 papers on technologies designed for people with VI, intending to facilitate and stimulate future research \cite{brule2020review}. Inan et al. surveyed 20 individuals who are visually impaired to investigate their internet use and explore their cybersecurity challenges and concerns while browsing the internet \cite{inanInternetUseCybersecurity2016}. They found several problems, such as automatic web page refreshing and missing or improper headings.
In this study, participants rated the possibility of someone tracking their internet activities as their highest-rated concern. The authors suggest that it is important to guide the user to enable security and privacy settings and to provide accessible software solutions to protect and warn this marginalised group.
Hayes et al. shadowed and interviewed people with VI and their allies \cite{hayesCooperativePrivacySecurity2019}. Finding that self-perceptions can influence someone's behaviour, which could have privacy and security implications, such as hiding or concealing their disability due to perceived stigma. Akter et al. studied 155 people with VIs privacy concerns relating to camera-based AT \cite{akter2020uncomfortable}.  
Finding that users of these systems were more concerned about the privacy of others, who may inadvertently be captured in their images, than themselves. However, camera-based AT can create a lack of personal security in the lives of the people they are trying to help.
Previous research reports that users with VIs often find it difficult to complete their security tasks and that they had moderately high levels of concern about cybersecurity \cite{napoli2021m}. Similarly, there are reports on the complications of authentication methods such as passwords and two-factor authentication for users with VIs \cite{schmeelk2022digital}. An exploratory user study, conducted using semi-structured in-person interviews with 14 people with VIs found that participants were aware and concerned about privacy and security and faced a variety of risks \cite{ahmedPrivacyConcernsBehaviors2015}.

More relevant to this paper is the work of Schnell and Roy~\cite{schnellWebsitePrivacyNotification2021}. They evaluated a select group of 40 educational and financial website cookie notices using WCAG. Finding that even for users without disabilities, there were challenges to accessing, understanding, and processing privacy information. Also, found that educational websites were more accessible than financial websites, however, not all websites complied with the WCAG's criteria chosen for their testing. In contrast to this work, we offer a comprehensive evaluation method to review website cookie notices and apply our methods to a range of websites rather than only educational and financial. 

Although there have been several user studies looking generally at security and users who have a VI, to the best of our knowledge, there have been none looking specifically at cookie notices. In this paper, we aim to address this gap via a series of system studies and a dedicated user study with users who have VIs, both focusing on cookie notices.

\subsection{AT Regulations, Standards, and Tools}
According to GDPR, cookie notices should be presented on all websites that use cookies and should include opt-out options, as well as opt-in options without highlighting the latter and including any privacy or cookie walls. Cookie notices should be separated from other matters such as privacy policy and terms and conditions, and the user should be able to opt out of the previously accepted cookie settings with the same ease as they gave the consent. Enabling non-essential cookies before the user's consent is a non-compliant practice too. 
Based on Article 12 et seq. GDPR~\cite{gerl2019layered}: ``The controller shall take appropriate measures to provide any information referred to in Articles 13 and 14 and any communication under Articles 15 to 22 and 34 relating to processing to the data subject in a concise, transparent, intelligible, and easily accessible form, using clear and plain language, in particular for any information specifically addressed to a child. The information shall be provided in writing, or by other means, including, where appropriate, by electronic means. When requested by the data subject, the information may be provided orally, provided that the identity of the data subject is proven by other means.'' This article interprets that the data controller~(i.e. web tracker in the context of this paper), must inform every user about the nature of the data to be collected and the purposes of such collection. Hence, websites need to be fully compliant with the regulations and also offer usable practices to comply with further requirements. 
The ambiguity of how practises should include marginalised users has not been discussed widely, and only limited examples are available. For example, in the verdict of an Italian case in which the data controller was mandated to provide the information acoustically for video surveillance \cite{italianCase}.

There are many aspects to the real-world implementation of accessible web technologies~\cite{pernice2001usability}. For instance, an accessible web design approach should support enhancing the visual characteristics of the front-end design and utilise a range of colours, while ensuring the contrast of the colours is accessible to users who are visually impaired or colour--blind. Also, they need to build an audio commentary for the page and the images. The interconnected nature of web pages~(as various resources fetched from different origins in the page) could potentially increase the complexity of fully accessible web design. To harmonise such practises, the W3C has provided a comprehensive list of 167 tools to evaluate accessibility compatibility measurements\footnote{W3C Web Accessibility Evaluation Tools List, \url{https://w3.org/WAI/ER/tools/}}. They are implemented on several platforms and technologies, some supporting cross-platform products. These products include 20 support APIs, 14 authoring tool plugins, 45 browser plugins, 19 command line tools, 25 desktop applications, 4 mobile applications, and 90 online tools.

There are a number of standards and regulations worldwide to provide accessibility requirements for the technologies to be considered as \emph{publicly presentable} in a region, country, or regional-based regulations e.g., European, Italian, Irish, Israeli, Japanese, Korean, and US Federal Law, platform~(web accessibility frameworks), e.g., various versions of WCAG~(2.1, 2.0 and 1.0), or file formats, e.g., EPUB standards. 

Looking at the standards based on the support for VI, we can conclude that all of them recognise such disabilities and provide a standardised set of guidelines for the implementation of such support. In general, they offer similar suggestions to aid users with such disabilities. For example, these standards support VI by advising to provide a form of AT and non-visual access to support VI users, including the proprietary agent~(in this case, the dedicated hardware or special browsers), an audio description to explain the important visual detail, the high contrast visualisation, adoption of flash thresholds, magnification, reduction of the required field of vision, and control of contrast, brightness, and intensity. Following these guidelines can contribute towards meeting the minimum requirements for complying with such regulations.  

There are various automated accessibility testing tools which can test the accessibility of a website~\cite{w3WCAGOverview2022}.
In our observations, most accessibility tools are based on W3C standard family~(WCAG 2.1~(85 tools out of 167), WCAG 2.0~(139 tools), WCAG 1.0~(46 tools)). Moreover, some of them comply with country-specific regulations such as German standards~(21 tools), French standards~(12 tools), Japanese standards~(18 tools), EU standards~(9 tools), US federal procurement~(67 tools), Irish National IT Accessibility Guidelines~(16 tools), Israeli web accessibility guidelines~(7 tools), Italian accessibility legislation~(11 tools), Korean standards~(1 tool). Finally, format-specific standards such as EPUB accessibility 1.0 are only supported in 3 tools.

\subsection{Online Tracking} 

Adopted in April 2016 and implemented in May 2018, the GDPR changed the rules on online tracking and consent (including consent to cookies) \cite{GeneralDataProtection,matte2020cookie}. 
In order to process personal data, companies must choose a legal basis for processing.
One of the most well-known is consent. 
Valid consent must be freely given, specific, informed, unambiguous, explicit, revocable, given before any data collection, and requested in a readable and accessible manner \cite{GeneralDataProtection}.
The ePrivacy Directive (``ePD'', aka ``cookie law'') \cite{Directive2002582002}, provides supplementary rules to the GDPR. 
According to the ePD website, publishers must rely on user consent when collecting and processing personal data using non-mandatory (not strictly necessary for the services requested by the user) cookies or other technologies \cite{Directive2002582002}. This is in accordance with the guidance given by the European Data Protection Board and the ICO \cite{europeandataprotectionboardGuidelinesConsentRegulation2018,informationcommissionersofficeGuidanceUseCookies2022}. Various studies (e.g., \cite{utzInformedConsentStudying2019,degelingWeValueYour2019, mehrnezhadCrossPlatformEvaluationPrivacy2020,trevisanYearsEUCookie2019,mehrnezhadHowCanWould2022, huCharacterisingThirdParty2019}) exist on the implementation and effectiveness of cookie notices, privacy banners, and tracking practises. 
Examples of dark patterns include providing invalid consent, nudging the user, 
making opting-out difficult, not providing the user with opting-out options from previously accepted cookie settings, pre-enabling non-essential cookies, 
and including trackers in the cookie notice itself. For example, the top 5 consent management platforms have been reported to use dark patterns and implied consent \cite{nouwensDarkPatternsGDPR2020}.

There is a body of knowledge on the user dimensions of tracking, including concerns and negative feelings of users about tracking \cite{raoWhatTheyKnow2015}, differences between demographics such as gender and country \cite{coopamootooFeelInvadedAnnoyed}, the disparity between regulations, website practises and users' limited knowledge for protecting against tracking and their demand for more transparency and control \cite{mehrnezhadHowCanWould2022,shiraziWhatDetersJane2014,puglieseLongTermObservationBrowser2020,melicherPreferencesWebTracking2016,urSmartUsefulScary2012}. What is lacking in the previous work is the measurement of the current practises in the wild for web tracking notices for users with VIs. In this paper, we aim to run experiments in order to fill this gap. 

\section{Accessibility Evaluation Methodology}
In this section, we present our methodology for the evaluation of the websites. 

Our assessment includes a number of different methods and tools including automated accessibility testing tools, a non-graphical browser and screen readers, as explained at length in this section. The overall design of our experiments and the tools used in each part are presented in Table \ref{tab:overall}. We have included a website analysis template in Appendix \ref{website_analysis_template}.

All experiments took place between April and October 2022 on a laptop PC running Windows with a screen size of 13.3 inches and a resolution of 3840 x 2160. Windows is the most commonly used desktop OS among screen reader users according to the WebAIM 2021 survey \cite{webaimWebAIMScreenReader2021}. 
As a case study, we used Alexa's top 50 UK websites in April 2022.
We selected this sample since GDPR is a regional regulation (EU/UK). We looked at English websites for analysis in our fluent language. Based on Alexa, the popular UK websites are comparable to others in Europe, e.g., Germany and France.
From this list, four websites are excluded because they redirect to another website already on the list or are down. \emph{t.co} is an example of a website that was excluded due to redirecting to \emph{twitter.com}, however, both \emph{amazon.com} and \emph{amazon.co.uk} are retained.
The US version of the site (.com) does not contain a cookie notice, whereas the UK version (.co.uk) does; therefore, it was important to keep both sites on the list for comparison. These are just examples, and the full list is presented in Table \ref{tab7} (Appendix). The cookie notice experiments were conducted by two researchers to ensure consistency. A researcher performed accessibility testing twice with the specialist software/tools, recording the results in tables (Appendix). 
Due to the rounds of experiments taking place over the course of six months, we believe this demonstrates stability in our results.

\subsection{Cookie Notices}
All 46 websites were visited using Google Chrome (Version 103.0.5060.134 (Official Build) (64-bit)) and Brave\footnote{Brave, \url{https://brave.com}} (Version 1.41.99 Chromium: 103.0.5060.134 (Official Build) (64-bit)). Using Google Chrome without a screen reader acts as a baseline and gives an example of how sighted users would see the site and the cookie notice. Chrome is one of the most popular browsers with the highest market share in 2022 \cite{BrowserMarketShare}. Brave is a secure browser that was created in 2016 by two former executives of Mozilla Corporation, the company that makes the Firefox browser \cite{lundBraveBrowserMonetary2021}. Brave comes with a feature called Brave Shields built in, which includes several privacy-preserving features. Brave adopts various privacy-enhancing techniques
which are not possible at the browser extension level (due
to access restrictions and performance limitations), making it a powerful tool for observing the tracking behaviours of
websites. It is commonly used for assessing the tracking behaviour of websites on PC and mobile platforms \cite{mehrnezhadCrossPlatformEvaluationPrivacy2020,mehrnezhadHowCanWould2022}. We completed these experiments before the introduction of cookie notice blocking by Brave \cite{thebraveprivacyteamBlockingAnnoyingPrivacyharming2022}. 

For each of the 46 websites, we open them in these two browsers and record the location and control options given to users. 
When recording the details, we do not interact with the website in any way, including not interacting with notifications (e.g. requesting location permission or update notifications).
To ensure that no cookies had previously been cached, each website is viewed in a new private or incognito window.

We also record which options are given to the user in the cookie notice according to the categories suggested in similar work, e.g. \cite{mehrnezhadCrossPlatformEvaluationPrivacy2020, mehrnezhadHowCanWould2022}. 
These categories include: 
(i) \emph{Agree or Reject}: where two options are presented, Agree (Agree, Accept, OK, Understand, etc.) or Reject (Reject, Decline, No, etc.), with the same level of control (e.g., two buttons). These are further categorised by which option is emphasised.
(ii) \emph{Agree or Settings}: where two options are presented, Agree or Settings (Options, Settings, Policy, Manage, Learn more, etc.), again with the same level of control. Which are further categorised by which option is emphasised.
(iii) \emph{Agree, Reject, or Settings}: where three options are presented; Agree, Reject, and Settings. These are further categorised based on which item is highlighted in the notice. 
(iv) \emph{No Notice}: The website does not display a cookie notice.

\subsection{Accessibility Evaluation} 
Our accessibility evaluation consists of two parts. First, we use automated testing tools, which are designed to give developers an overview of how accessible their website is \cite{w3WCAGOverview2022}. This allows us to get an impression of the overall accessibility of a website, and in some cases the accessibility of the cookie notice. Second, we use software designed for individuals with VI in the real world to assess the results of the automated testing tools and to allow us to focus more specifically on cookie notices. In this section, we explain these approaches. 

\textbf{Automated Accessibility Testing Tools:} 
Websites are evaluated using two different automated accessibility testing tools, WebAIM WAVE 5.0 Web Accessibility Evaluation Tool\footnote{WAVE, \url{https://wave.webaim.org/}} and Google Lighthouse\footnote{Google Lighthouse, \url{https://developer.chrome.com/docs/lighthouse/overview/}}. 

WAVE is an automated accessibility tool that we use to perform an initial assessment of the conformance of each website to WCAG. WAVE generates a report containing Errors, Alerts, Features, Structural elements, and ARIA landmarks. 
Errors indicate issues that will impact certain users with disabilities, as well as showing failures to meet the requirements of the WCAG. Whereas alerts are elements which may cause accessibility issues but need further manual testing to determine this.
Features are elements that can improve accessibility if implemented correctly.
Structural elements are HTML regions and headings, and ARIA can be used to present important accessibility information to people with disabilities.
WAVE has been used in previous work, e.g. Southwell and Slater, when evaluating university library finding aids \cite{southwellEvaluationFindingAid2013}. During their testing, they used WAVE to perform an initial evaluation of the conformity of each finding aid to Section 508 and WCAG 2.0 guidelines. We tested the web version of WAVE 5.0 in our preliminary testing and it did not detect any cookie notices. Therefore, we use the browser extension version\footnote{WAVE browser extension, \url{https://wave.webaim.org/extension/}} for our experiments.
The WAVE extension evaluates the rendered version of the web page allowing dynamically generated content to be evaluated \cite{WAVEExtension}, while the WAVE Web version may not be able to apply all the scripting on the page. This is a possible reason for the cookie notices not being displayed during our preliminary tests.

We use Google Lighthouse to give an overall accessibility score, as well as to record specific problems with each website. Lighthouse is an open-source automated testing tool, which can audit performance, accessibility, and more \cite{googledevelopersLighthouseOverview2022}. We only test accessibility using the default (navigation) mode and while representing a desktop device. We record the score out of 100 and the individual issues with each website. 
This score is a weighted average of all accessibility audits it performs, with weighting based on axe user impact assessments \cite{LighthouseAccessibilityScoring}. The axe user impact assessments are composed of WCAG rules with some supplementary rules added \cite{Axecore2022}.

Both WAVE and Google Lighthouse give an overview of accessibility for a whole website, however, WAVE also allowed us to view where specific problems occurred.

\textbf{Manual Testing via Text-only Browser:} 
To complement and verify the results of the testing tools, we apply a range of methods to manually assess the privacy practises of these websites via their cookies notices. 
We visit all these websites using WebbIE\footnote{WebbIE, \url{https://webbie.org.uk/}}, a text-only browser for people with VI. The WebbIE Ctrl-H and Ctrl-L commands are used to examine the heading and links on a page. This approach has been used in similar work, e.g., work \cite{southwellEvaluationFindingAid2013}. WebbIE was uninstalled and reinstalled for each round of testing, as it does not have a private browsing mode or cookie manager. Through this method, we examine how users navigate the page as well as if and how cookie notices are displayed. We assign each website to one of the following categories: 
\emph{(i) No Headings}: The website in general has no headings which can be used for navigation.
\emph{(ii) Basic Headings}: The website has some headings but there is a limited amount which is not useful for navigation. 
\emph{(iii) Full Headings}: The website has a number of headings that are useful for navigation.

Headings allow screen readers and other accessibility software to navigate around a webpage. For example, WebbIE can move easily to different headings on a website allowing for quicker navigation and locating key information, e.g. a cookie notice. 
The categories above are derived from previous work \cite{southwellEvaluationFindingAid2013}, where similar categories were used to evaluate the accessibility of library finding aids. Similarly, we observed the website's behaviour in presenting the cookie notice and each website's cookie notice was also put into the following categories:
\emph{(i) Headings throughout}: Headings are available throughout the cookie notice.
\emph{(ii) Heading at the start}: A heading is present at the start of the notice, however, there were no other headings in the body of the notice.
\emph{(iii) No headings}: There are no headings present in the cookie notice at all.
\emph{(iv) Notice missing}: The cookie notice is not shown when using WebbIE, however, one is present when using the graphical browsers.
\emph{(v) No notice}: The website has no cookie notice when viewed with the graphical browser.

The \emph{Headings throughout} category for cookie notices is based on the \emph{Full Headings} category for the website as a whole. This means that a user would be able to navigate the cookie notice using heading-based navigation; this is particularly useful for longer cookie notices as seen on some websites.
The \emph{Heading at the start} category is used to classify notices that only have a heading at the start. This would allow for navigation to the notice itself but means that a user would have to rely on a different type of navigation within the notice, e.g. line-by-line or link-based navigation.
Whereas \emph{No headings} would mean a user would not be able to use heading navigation at all within the cookie notice and would have to rely on another form of navigation.
In some instances when viewed graphically, a website did display a cookie notice, however, when using WebbIE one was not present. For this reason, we include a \emph{Notice missing} category to signify this. 
Whereas with the \emph{No notice}, a notice was not present on the website when viewed graphically. 
We included different categories for headings (Basic, Full, and No headings) since we found lengthy cookie notices on some websites (e.g., google.com, facebook.com), however, headings are not always needed due to several cookie notices being shorter.

\textbf{Manual Testing via Screen Readers:} 
In order to have more comprehensive and conclusive results, we also carry out our experiments using screen readers to test each website manually. JAWS and NVDA were chosen as they are the most popular according to WebAIM~\cite{webaimWebAIMScreenReader2021}, 53.7\% and 30.7\%, respectively. We use these screen readers in conjunction with Google Chrome as these are the most common combinations of screen reader and browser~\cite{webaimWebAIMScreenReader2021}, 32.5\% and 16.0\%.
NVDA is a free OS-level screen reader with support for popular applications such as web browsers, email clients, music players, and office programs. JAWS is another OS-level screen reader that users need to purchase. For our experiments, we purchased a home licence (£865 with the Authorisation USB Dongle). Both screen readers should have similar reliability when parsing websites \cite{webaimScreenReadersCSS2017}, however, they often parse website code slightly differently \cite{powermappersoftwareScreenReaderReliability2022}. It is for this reason that we use the two most popular screen readers during our testing.

We categorise each website's cookie notice based on how these screen readers can parse them~\cite{w3HowMeetWCAG2019,schnellWebsitePrivacyNotification2021}. Accordingly, each website’s cookie notice was given a pass or fail for the following categories:  
\emph{(i) Readable}: The screen reader software is able to read the cookie notice. 
\emph{(ii) Immediately Read}: The cookie notice is the first thing to be read from the page, excluding the page title.
\emph{(iii) Keyboard navigable}: The cookie notice of a website is navigable using a keyboard while using a screen reader.
\emph{(iv) Link or button purpose}: The purpose of a link or button can be solely determined by the link or button. 
\emph{(v) Abbreviations are explained}: All abbreviations are explained. This was either in the cookie notice or the website offered some mechanism for identifying the expanded form. 
\emph{(vi) Page titled}: The page has a title that describes its topic or purpose.
\emph{(vii) Cookie notice titled}: The cookie notice has a title or heading which is readable by the screen reader software.
\emph{(vii) Headings useful for navigation}: There are useful headings for navigation present throughout the cookie notice.
\emph{(ix) No timing}: There is no timing for reading the cookie notice.

The \emph{Readable} category is based upon WCAG guideline 3.1, Readable, defined as ``Make text content readable and understandable'' by WCAG \cite{w3HowMeetWCAG2019}, with the guideline being used in previous work \cite{schnellWebsitePrivacyNotification2021}. 
We created the \emph{Immediately Read} category to show that a cookie notice is read close to the start of a web page. 
This is important as a number of websites start tracking a user before they respond to the notice, and therefore users must be able to respond to the notice at the first given opportunity. Also, this means that users do not have to actively search the website for the notice to respond. 
The category \emph{Link or button purpose} is based on WCAG guideline 2.4.9, Link purpose (link only), which is defined as ``A mechanism is available to allow the purpose of each link to be identified from the link text alone, except where the purpose of the link would be ambiguous to users in general'' \cite{w3HowMeetWCAG2019}.
\emph{Abbreviations are explained} is based upon WCAG guideline 3.1, Abbreviations, which W3C define as ``A mechanism for identifying the expanded form or meaning of abbreviations is available''. 
\emph{Page titled} is also based on WCAG, namely guideline 2.4.2 Page Titled which is defined as ``Web pages have titles that describe topic or purpose''. 
We create another category based on this called \emph{ Cookie notice titled}, this is to judge if a cookie notice can easily be navigated to. It also aligns with previous testing for headings, as titles often consist of headings. 
Alongside this, we test for \emph{Heading useful for navigation}, which is based on the previous heading testing with WebbIE in cookie notices. 
It also aligns with WCAG guideline 2.4.10 Section headings, defined as ``Section headings are used to organize the content''. 
We also define the category \emph{No timing} which is based on WCAG guideline 2.2.3 No timing. 
This is defined as ``Timing is not an essential part of the event or activity presented by the content, except for non-interactive synchronized media and real-time events''.

\subsection{Limitations}
To the best of our knowledge, this is the first work on the assessment of cookie notices on a range of websites for users with VI. We simply chose to test the 46 top websites in the UK (out of 50). In practice, the top Alexa websites may not be the most popular websites for users with VI. However, we could not find a formal report on popular websites for this group of users. We acknowledge that this is a limited sample set and more research is required to evaluate a larger number of websites. 
When testing websites, we only tested the first page in our experiments. Although this is a common practice for the privacy assessment of websites in general, it is not clear if all pages would present the same information and produce the same output for AT. Further, detailed work would be needed to explain how different web pages interact with AT.

Previous research has demonstrated the usefulness of mobile technology for people with VI, e.g. \cite{griffin2017survey, hakobyan2013mobile}. However, due to the lack of research in this area, we only generally focus on desktop web browsers, for which the majority of the accessibility and AT tools and standards are also designed. Cross-platform studies are left as future work. 

\section{Accessibility evaluation Results}
Our results include (1) a general assessment of the cookie notices of the websites and their tracking behaviour, and (2) an accessibility evaluation of these websites and their cookie notices. 

\subsection{Cookie Notices and Tracking Behaviour}
\textbf{Cookie Notice Position:} 
We observed that the majority of websites displayed a cookie notice (n=35 or 76.1\%) when using Google Chrome. Of the positions, a bottom overlay was the most common (n=15 or 32.6\%), followed by a middle overlay (n=7 or 15.2\%).
When using Brave, a higher number of web pages displayed no notice (n=15 or 32.6\%). Other than this, the popularity of categories is in the same order as that of Google Chrome. While there are some papers (e.g. \cite{utzInformedConsentStudying2019, bermejo2021website}) looking at cookie notice positions and user engagement, we could not find any for users with VI.

\textbf{Cookie Notice Control Options:} Of the options given when using Google Chrome, Agree or Settings was the most common (n=17 or 37.0\%). The most commonly emphasised option along with Agree or Settings was Agree (n=13 or 28.3\%). Table \ref{tab:options} describes the options presented to users in Chrome and Brave. These results from Brave resemble those of Google Chrome; however, when using Brave, there was a higher percentage of websites which displayed no notice.
These results are consistent with previous work (e.g., \cite{mehrnezhadCrossPlatformEvaluationPrivacy2020}) when cookie notices were evaluated across platforms. The cookie notices of some websites are blocked due to the notice itself being a tracker; resulting in it being blocked by Brave. 

Previous research studying GDPR compliance has focused on the following requirements: consent must be explicit, accepting all is as easy as rejecting all, and there are no pre-ticked boxes \cite{nouwensDarkPatternsGDPR2020}. It has been shown that the pre-selection of options can impact users' choices when giving consent \cite{utzInformedConsentStudying2019}. For this reason and to respond to RQ1-a, in Table \ref{tab:options}, we highlight categories that violate the above requirements and therefore in violation of GDPR. As you can see, three categories (14 websites) comply with the above requirements. However, we did not test them for additional GDPR compliance items, such as opting out from previously accepted cookie notices with the same ease of opting in. 

\begin{table}[t]
    \centering
    \footnotesize
    \caption{Cookie notices' user control options in Chrome and Brave, as well as GDPR violations.}
    \label{tab:options}
    \begin{tabular}{llrrr} \toprule
        \textbf{Control} & \textbf{Emphasised}&\multicolumn{2}{c}{\textbf{Browser}} &\textbf{GDPR}\\ \cmidrule(r){3-4}
         \textbf{options} &  \textbf{option} & \textbf{Chrome} & \textbf{Brave} &  \textbf{violation}\\ \midrule
       (i) Agree  or Reject & None   & 4 & 4 & No\\
        & Agree   & 4 & 4 &Yes\\
      (ii)  Agree  & None   & 4 & 4 & Yes\\
      or Settings  & Agree   & 13 & 9 & Yes\\
     (iii)   Agree, Reject  & None & 5 & 5 & No\\
      or Settings & Agree \& Reject & 5 &5 & No\\
        \multicolumn{2}{l}{ (iv) No Notice} & 11 & 15 &Yes\\ \bottomrule
    \end{tabular}
\end{table}

\textbf{Tracking Behaviour:} We also observed these websites regarding their tracking behaviour through Brave. Before interacting with the cookie notices 
only 3 of the 46 websites (6.5\%) did not have at least one item blocked by the Brave Shields. The average number of items blocked was 9, the maximum was 81, 11 of the websites had more than 10 items blocked, and 6 had more than 20. The majority of items blocked were in the \emph{trackers \& ads} category. Our results support similar work e.g., \cite{mehrnezhadCrossPlatformEvaluationPrivacy2020, mehrnezhadHowCanWould2022,matte2020cookie}, reporting that the majority of websites start tracking the user regardless of the presence of the cookie notice before any user interaction with it.

\subsection{Automated Accessibility Testing Tools in Websites}
WAVE ran on all but one website (n=45); when using it on \emph{ebay.co.uk}, the overlay containing the results did not appear. Table \ref{tab4} shows a summary of the results of testing with WAVE. We further break these down into categories which could cause issues, e.g. errors, contrast errors, and alerts, and those which could improve user experience, e.g. features, structural elements, and ARIA.
Of the sites tested, 41 (91\%) contained at least one accessibility error, with the average number of errors being 19. Errors are general issues which may cause problems, for example missing HTML code, or missing alternative text.
Of the websites tested  34 (76\%) contained at least one contrast error. A contrast error would cause issues for someone with vision loss, e.g., light text on a light background or dark text on a dark background.
44 sites contained at least one alert, which needs further testing to establish if they hinder or help accessibility. 
For example, for an image with long alternative text, a long description could be needed to fully describe the image, or it may be unjustified.
All websites tested contained at least one structural element with the average being 84, all websites tested also contained at least one feature with the average number being 77.2. Features are elements which work to improve a user's experience. For example, a form label is present and associated with a form control. 
This is similar to structural elements such as headings and lists which also help the user's experience. 42 (93\%) of the tested sites contained at least one ARIA element. ARIA is a set of roles and attributes that define ways to make websites more accessible to people with VI. An example of an error could be missing alternative text or a form control that does not have a corresponding label. ARIA is only useful if implemented correctly such as when an `aria-label' or `aria-labelledby' attribute is present which can be interpreted by AT.

\begin{table}[t]
    \centering
    \footnotesize
    \caption{Summary of WAVE 5.0 test results for 45 websites.}
    \label{tab4}
    \begin{tabular}{l|rr} \toprule
        \multirow{3}{*}{\textbf{Criteria}} & \textbf{No. of websites}& \textbf{Average no.} \\ & \textbf{ with at least one }& \textbf{of items}\\
       &\textbf{item per criteria}& \textbf{across websites} \\\hline
        Cookie Notice & 33  & -\\\hline
        Errors & 41  & 19\\
        Contrast Errors & 34  & 22.9\\
        Alerts & 44  & 124\\\hline
        Features & 45  & 77.2 \\
        Structural Elements & 45  & 84\\
        ARIA & 42  & 235.4\\ \bottomrule
    \end{tabular}
\end{table}

To complement this, we also note whether a cookie notice was present when testing using WAVE. With 33 (73\%) cookie notices being present. In some instances, we observed specific issues with the cookie notice. The most common problems were the low contrast between the background of the cookie notice and the text, links, or buttons.
Overall, the website with the lowest number of issues was bbc.co.uk, with 0 errors, 0 contrast errors, 134 alerts, 23 features, 119 structural elements and 371 ARIA. There were multiple websites with close to the same number of issues, namely xvideos.com, spankbang.com and xnxx.com, all of which had between 165 and 176 items which would cause issues (i.e., errors, contrast errors, \& alerts). 

In addition, we used Google Lighthouse for the overall accessibility score of the website. The average score was 89\% (highest: 100\%, lowest: 63\%). Since Google Lighthouse uses the axe user impact assessments, the overall score is affected largely in the same way as individual WAVE tests. For example, including a button that has an accessible name will improve the overall score given to a web page. 
In general, these popular websites had a range of good and poor accessibility practices when tested with these automated accessibility tools. 
There were several websites we tested that achieved the best possible Lighthouse score of 100\%: bbc.co.uk, wikipedia.org, gov.uk, paypal.com, microsoft.com, linkedin.com, and doubleclick.net. The lowest score, 63\%, was achieved by tiktok.com. which, according to Lighthouse, had a number of labels and names missing, as well as navigation and contrast issues. Detailed results for each website, including both WAVE and Lighthouse tests, are shown in Table \ref{tab7} (Appendix).

\subsection{Manual Cookie Notice Accessibility Testing via Assistive Technology}
\textbf{Manual Testing via Text-only Browser}: By using a text-only browser, we performed an analysis of the overall accessibility of the websites and their cookie notices. 
When using WebbIE, 27 of the 46 (58.7\%) websites contained \emph{full headings} which would be useful for navigation. With 7 (15.2\%) of them only having \emph{basic headings} and 12 (26.0\%) containing \emph{no headings}. The inclusion of headings throughout the website does not directly impact privacy and was included in this analysis to give context to the accessibility of cookie notices.

\begin{figure}[t]
   \input{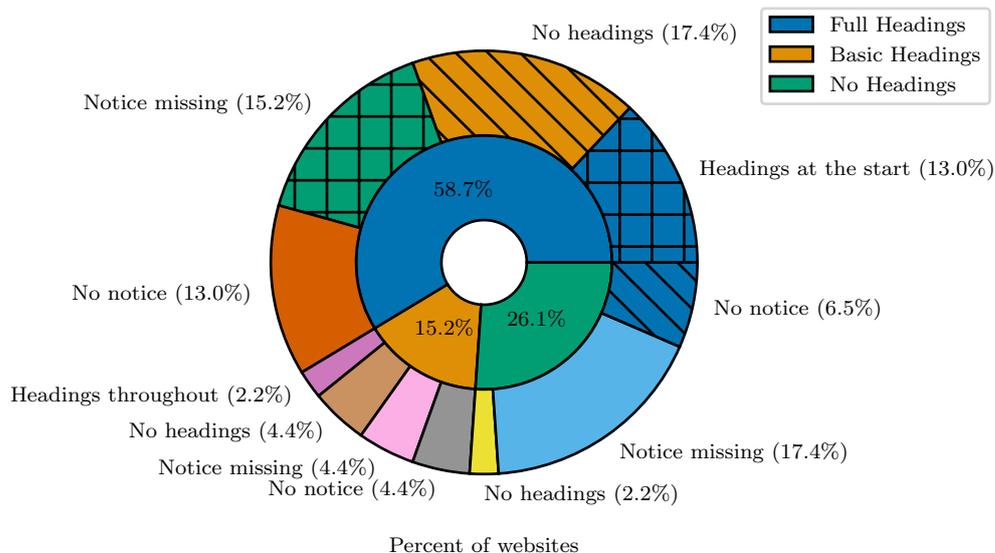}
   \caption{WebbIE accessibility testing; inner circle: the whole site, outer circle: the cookie notice.} 
   \label{fig4} 
   \alt{Sunburst chart showing results for manual WebbIE testing, with the inner circle representing the whole webpage and the outer representing the cookie notice. The values for the inner circle are Full Headings (58.70\%), Basic Headings (15.22\%), and None (26.09\%). Outer circle values for Full Headings: Headings at the start (13.04\%), No headings (17.39\%), Notice missing (15.22\%), No notice (13.04\%). Outer circle values for Basic headings: Headings throughout (2.17\%), No headings (4.35\%), Notice missing (4.35\%), No notice (4.35\%). Outer circle values for No headings: Notice missing (17.39\%), No headings (2.17\%), No notice (6.52\%).}
\end{figure}

When observing the cookie notices, 17 (48.6\%) of the 35 websites which previously displayed a cookie notice did not display a cookie notice when using WebbIE (\emph{notice missing}). 
Furthermore, only 1 (2.9\%) of the 35 websites which had previously displayed a cookie notice website had \emph{headings throughout}, and 6 (17.1\%) had a \emph{heading at the start} of the notice. 11 (31.4\%) of the websites' cookie notices contained \emph{no headings}, although, the majority of the websites which did not contain a notice did include links to privacy and cookies. 
Regardless of the number of headings throughout the website, we often found that cookie notices headings were missing. However, when a website had full headings the cookie notice was more likely to have a heading at the start.
The results of these tests are shown in Figure \ref{fig4}, with the inner circle representing the headings in the website as a whole and the outer circle specifically looking at the cookie notice.
The use of a heading at the start of the cookie notice can make it easier to locate, due to this the lack of headings seen in our testing could lead to problems. 
Headings inside the notice can also make it easier to navigate within a cookie notice, especially if it is lengthy, and therefore easier to make a decision.

\textbf{Manual Testing via Screen Readers}: When testing with NVDA, 29 (82.9\%) of the 35 websites, which graphically included a cookie notice, contained a cookie notice which could be read aloud. This result is higher than was expected following the other test. However, it still means that 6 of the cookie notices could not be read at all with NVDA. 
Of the cookie notices that could be read, 20 of the 35 (57.1\%) were read aloud immediately when the website loaded. Others were read aloud after other elements of the page had been read or had to be specifically located to be read. 27 (77.1\%) of the 35 cookie notices were keyboard navigable, these were not always the same websites as those read immediately. Therefore, this leaves 8 websites which users with VI may not be able to navigate. In some cases, these cookie notices created keyboard traps that the user would not be able to leave. Only 5 (14.3\%) of the 35 cookie notices contained a link or button that could be solely determined by the link or button. Hence, without allowing time for the screen reader to output the notice, the user may not understand what they are agreeing to. 

Although all 46 (100\%) websites contained a title, only 19 of the 35 (54.2\%) cookie notices contained a title. This means it would not be possible to navigate to them using the heading, it could also make it more difficult to search for the cookie notice. 35 of the 35 (100\%) cookie notices did not have any type of time limit on replying to the cookie notice. This is an excellent result, meaning that users will have time to ingest the information and make a decision. None of the 7 websites which contained abbreviations explained them, meaning that if users are unfamiliar with these terms they may not understand what they are consenting to. Also, none of the 35 websites' privacy policies contained headings which were useful for navigation, however, some did contain different links. Due to this, it may be difficult to navigate the cookie notices, which is particularly important for some of the longer cookie notices we observed. We summarise these results in Table \ref{tab5} (detailed results in Table \ref{tab11} (Appendix)).

In comparison, JAWS enabled 34 (97.1\%) of the 35 websites with a cookie notice to be read out loud. This means all but one of the cookie notices could be read aloud, which is a significantly better result than when using NVDA. 
Of these, 22 (62.9\%) of the 35 were read aloud immediately when the website loaded, which again is higher than when using NVDA. 29 of the 35 (82.9\%) were keyboard navigable, this is an improvement of 2 cookie notices over NVDA. The number of cookie notices with a link or button that could be solely determined by the link or button was also higher at 11 (31.4\%) of the 35. All of the other results were the same for JAWS as NVDA. 
These results are summarised in Table \ref{tab5} and detailed results are available in Table \ref{tab11} (Appendix).
The reason for the disparity in results is due to differences in how the screen readers parse webpages, resulting in differing numbers of readable notices. This underscores the importance (lack) of standardisation efforts.

\begin{table}[t]
    \centering
    \footnotesize
    \caption{Number of websites which passed and failed each criterion of the manual testings via NVDA and JAWS.} \label{tab5}   
    \begin{tabular}{lrrrr}\toprule
        \multirow{2}{*}{\textbf{Criteria}} & \multicolumn{2}{c}{\textbf{NVDA}} & \multicolumn{2}{c}{\textbf{JAWS}} \\  \cmidrule(r){2-3}\cmidrule(r){4-5}
         & \textbf{Pass} & \textbf{Fail} & \textbf{Pass} & \textbf{Fail} \\ \midrule
        Readable & 29 & 6 & 34 & 1 \\
        Immediately read & 20 & 15 & 22 & 13\\  
        Keyboard navigable & 27 & 8 & 29 & 6\\
        Link or button purpose & 5 & 30 & 11 & 24\\
        Abbreviations are explained & 0 & 7 & 0 & 7\\
        Page titled & 46 & 0 & 46 & 0\\
        Cookie notice titled & 19 & 16 & 19 & 16\\
        Headings useful for navigation & 0 & 35 & 2 & 33 \\
        No timing & 35 & 0 & 35 & 0\\ \bottomrule
    \end{tabular}
\end{table}

We identified poor practices on some of these websites. For instance, a news website (dailymail.co.uk) reads out adverts immediately before reading anything else such as the navigation bar or the cookie notice. This is highlighted in Figure \ref{fig:dailymail} (Appendix). This is even though this website's cookie notice is displayed using a large portion of the website. 
Another example was an online payment site (paypal.com), which read the body of the website aloud before reading the cookie notice. This aligns with the cookie notice being visually at the bottom of the page; however, this means that a user with VI using a screen reader could easily miss the cookie notice. An example of the visual representation of this notice and a scripted output of the website while using JAWS is available in Figure \ref{fig:paypal} (Appendix). We highlight this example, however, a similar output was common across multiple websites.
One social news website (reddit.com) was the only website with a cookie notice which could not be read with either screen reader, even with intervention with mouse input. Visually the cookie notice was located at the bottom of the window, however, it could not be selected with the screen readers. A visual example of the cookie notice is included in Figure \ref{fig:reddit} (Appendix).
In contrast, some of the websites presented the user with reasonable options when using a screen reader. For instance, bbc.co.uk clearly presented the users with opt-in and settings options. A scripted version of such output via NVDA screen reader is provided in Figure \ref{fig:bbc} (Appendix).

\section{User Study Methodology}\label{ss:user_studies}
In this Section, we explain the design of our online survey, data collection, and analysis.

\subsection{Questionnaire Design}
When designing this survey, we followed the design principles for questionnaires for people with VIs \cite{kaczmirekSurveyDesignVisually2007}.
Specifically, we aimed to inform participants about the topic of the survey before beginning the questionnaire, we indicated the type of answer after each question, and we also started each question with a consecutive number. We conducted an accessibility evaluation before conducting the survey, using the tools mentioned above, during which we did not find any issues.

Our questionnaire is made up of five sections---Internet and AT, Privacy-enhancing technology usage, Cookie notices, Suggestions, and Demographics---with the complete questionnaire included in Appendix \ref{questionnaire}.

\textit{Internet and AT:}
After verifying the screening questions, we asked our participants a few background questions about technology usage, such as what devices and AT they use. We list different AT technologies including screen readers, braille displays, text-only browsers, magnification software, and assistive browser extensions based on our research as well as allowing participants to add additional items.

\textit{Privacy-enhancing technology usage:}
Next, after a brief explanation of privacy-enhancing technologies (PETs), we ask participants which PETs they use listening to them according to the categorisation suggested in \cite{coopamootooFeelInvadedAnnoyed}, where the authors measure the correlation of people's feelings about tracking and their protective actions. 
These categories include: browsers do not track, virtual private networks, private browsing, password managers, privacy-focused web browsers, encrypted messaging apps, ad blockers, and file encryption tools, we additionally allowed participants to name other tools.

\textit{Cookie notices:}
We also ask our participants what they think cookie notices are and what they think they are supposed to do. As well as how they feel about cookie notices and how they interact with them. For the design of these questions, we followed \cite{mehrnezhadHowCanWould2022,coopamootooFeelInvadedAnnoyed}. 
We ended this section by asking if they have encountered issues with cookie notices, what they were, and why they think they happened.

\textit{Suggestions:}
Finally, informed by the results of our website experiments, we ask participants who they believe should be responsible for ensuring the privacy and accessibility of websites. We also ask which of our suggestions would help improve their experience online.

\textit{Demographics:}
We conclude by asking demographic questions. 

\subsection{Data Collection and Analysis}
We conducted our user studies via Prolific Academic\footnote{Prolific Academic, \url{https://prolific.com/}} among UK participants.
We conducted one initial testing round of the survey with 10 participants, asking for feedback upon completion. 
We fixed minor typos and made a few structural changes accordingly. At this stage and throughout further participation we received no complaints relating to the accessibility of the questionnaire.
We then distributed the questionnaire to a further 100 participants. 
We chose to use Prolific Academic to distribute the survey as this user group is notoriously difficult to recruit, therefore using a paid platform allowed us to gain this sample size. 
We rewarded participants at a rate of £12 per hour, which is categorised as `Great' by Prolific Academic.
This research received full approval from the University of Surrey's Ethical Committee before the research commenced.

Our method for processing the collected data is a mix of quantitative and qualitative analysis. 
For our free-text style questions, we run thematic analysis \cite{coopamootoo2017privacy}; taking an inductive approach and allowing the data to determine our themes. We are confident that adopting a deductive approach would have yielded comparable themes.
Two researchers conducted the thematic analysis independently, and due to the small sample size all authors discussed and agreed on these themes.
Our research focused on exploring potential differences between users with visual impairments and those of previous work, allowing data to determine themes.
For lengthy and complex responses multiple themes were assigned to them.
We also chose participant responses which represent themes for inclusion in the paper.

\subsection{Limitations}
The interaction and intersection of online services and AT could be a complex research topic to be investigated by user studies. To complement the technical findings of this paper, we ran our studies on an online platform and through a survey that provided us with self-reported responses which has its own limitations. 
We acknowledge that our choice of platform and user study method might have introduced some bias in our results, since what people say might be different from what they actually do. 
For example, using an online survey could cause response bias, or the length of the survey could have caused survey fatigue, or using a paid platform could have caused incentive bias~\cite{MingAccept2021}. However, due to the fact that our accessibility evaluation result aligned well with our user study results, we believe these biases did not significantly influence our results. To further minimise these biases in future research, we
plan to extend our user studies to one-to-one interviews as well as focus groups to gain a deeper understanding of the implications of this user group. 

\section{User Study Results}
In this section, we present our findings of the user study. 
Our study was completed by 100 participants who self-certify as using AT and live in the UK. Our participants occupy different jobs ranging from students, educators, healthcare and social assistance to business, hospitality, and some not working. 59 participants identified as male, 38 as female, 2 as non-binary, and 1 chose not to disclose their gender. The age of our participants varied with the highest number of participants being part of the 25 to 34 group (n=33), followed by 45 to 54 (n=20), and 18 to 24 (n=18), a full breakdown of participants' ages can be found in Table \ref{tab:demographics} (Appendix).

\subsection{AT and PETs Usage}

Half of the users surveyed use magnification software, 42 use a screen reader, 22 use an assistive browser extension, 9 do not use any AT while browsing the web, and other forms of AT have less than 15 users. The majority of our participants had either basic (n=33) or moderate (n=29) experience with screen readers.
Participants use various online services, with the most popular being email (n=98), shopping and e-commerce (n=93), and social media (n=90).
Participants use various devices to access these services including personal computers (n=94), mobile phones (n=94), tablet computers (n=47), as well as various Internet of Things devices.
Table \ref{tab:demographics} (Appendix) gives an overview of the demographic questions. 
    
\begin{figure}[h]
        \begin{center}
            \input{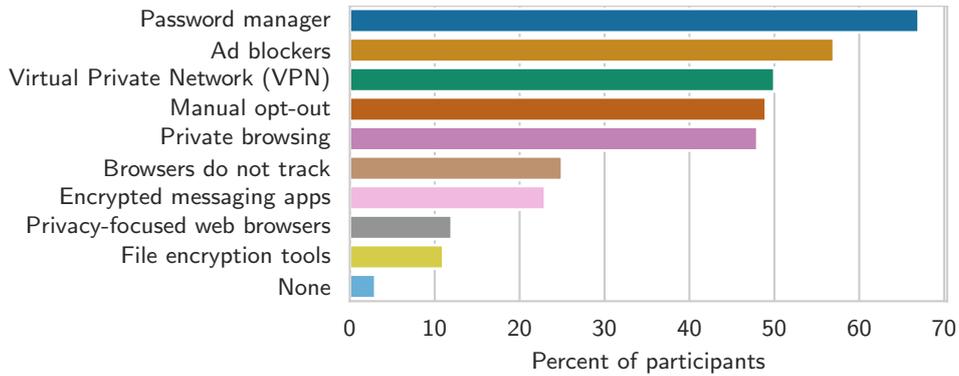}
        \end{center}

    \caption{Q2.1: Which of the following privacy-enhancing technologies do you use? (multiple choice)}
    \label{fig:which_pet}
    \alt{Bar chart showing the results for Q2.1 The results are: Password manager: 67, Ad blockers: 57, Virtual Private Network (VPN): 50, Manual opt-out: 49, Private browsing: 48, Browsers do not track: 25, Encrypted messaging apps: 23, Privacy-focused web browsers: 12, File encryption tools: 11, None: 3 }
\end{figure}

In response to Q2.1, which of the following privacy-enhancing technologies do you use, all but 3 participants use at least one of the PET categories suggested, Figure \ref{fig:which_pet} shows detailed results. The most popular technology used was a password manager (n=67), and the least popular was file encryption tools (n=11). 

We also asked about the ways these users learn about PETs. Participants reported different ways including recommendations of friends/social contacts, being informed at work/school, and news. Only 19\% of participants said they learned about these PETs via the privacy/cookie policy of a website, with the most common method being friend/social contact recommendation (n=46).

\subsection{Cookie Notices}

When we asked our participants about their understanding of a cookie notice (Q3.1), they described it via different terms and we observed a few themes, where one-third of our participants mentioned `tracking' with a negative tone. 
For instance, 
P94 said: ``It is a pop-up that appears on virtually every website I visit these days. Can be quite annoying since it collects data, but I tend to reject the tracking cookies if possible''.
We also observed themes around data(n=49), informing (n=47) and giving users options (n=25).

In response to Q3.2 on the feelings of the participants about cookie notices (Table \ref{tab:feel}).
Around half of our participants expressed negative feelings, one-third had neutral feelings, and around a quarter expressed positive feelings regarding cookie notices.
For instance, P9 said: ``I don't have any specific feeling about them just something that's there.'' and P33 said ``I don't like them, they are made difficult to understand on purpose, in order to make the user click "Accept". They need to be made more simple.'' 

\begin{table}[h]
    \centering
    \caption{Q3.2: How do you feel about cookie notices? }
    \footnotesize
    \begin{tabular}{llc} \toprule
        Category & Examples & \textit{N} \\\midrule
        \multirow{2}{*}{Strongly negative} & Don't trust, Intrusive & \multirow{2}{*}{24} \\
            & Very bad, Frustrating \\
        \multirow{2}{*}{Negative} & Dislike, Don't understand & \multirow{2}{*}{19} \\ 
            & Confusing \\
        \multirow{2}{*}{Neutral} & Okay, Not bothered & \multirow{2}{*}{31} \\
            &  Don't care \\
        \multirow{2}{*}{Positive} & Important, Essential & \multirow{2}{*}{26}\\ 
            & Useful \\
        \bottomrule
    \end{tabular}
    \label{tab:feel}
\end{table}

In Q3.3 and Q3.4, we asked the participants how they interact with cookie notices (Table \ref{tab:interact}). 
The responses varied across categories including agree (n=47), decline (n=34), ignore (n=8), edit cookie settings (n=7), get rid of it (n=7), and use other PETs (n=6).
Except for those who said they would agree to the cookie notice (47\%), all the other categories included the word ``try'' in some of the responses e.g., ``try to decline'' and ``try to edit the settings and say no.''
Interestingly, 7\% of participants spoke about trying to get rid of the cookie notice in any possible way by e.g., responding quickly. P46 said: ``generally tick as little as possible to view the page and also reject where I can if not[,] I have to accept if the page [is needed]''. Whereas P13 said ``I try to reject them but this can be very difficult- I find they are often deliberately set up to make it impossible to read.'' 

\begin{figure}[!ht]
    \input{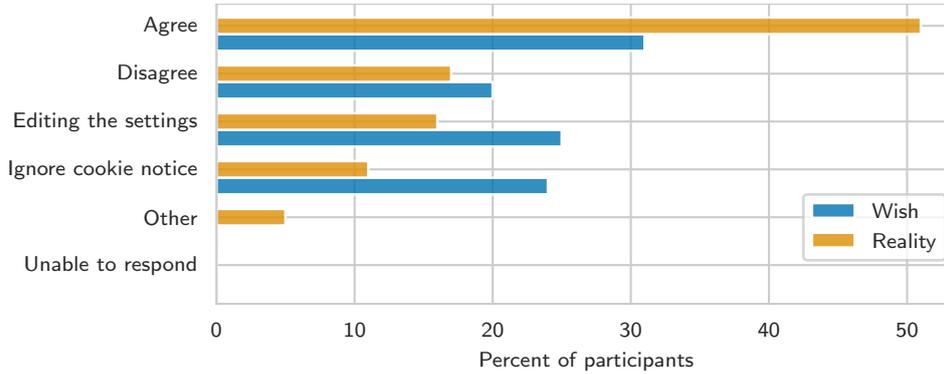}
    \caption{Q3.7: How would you like to handle cookie notices? and Q3.8: How do you actually respond to cookie notices?}
    \label{fig:How do people wish or really interact with cookie notices}
    \alt{Bar charts showing the answers for Q3.7 and Q3.8. The results for Q3.7 are: Unable to respond: 0, Disagree: 20, Ignore cookie notice: 24, Editing the settings: 25, Agree: 31. The results for Q3.8 are: Unable to respond: 0, Other: 5, Ignore cookie notice: 11, Editing the settings: 16, Disagree: 17, Agree: 51
    }
\end{figure}

In response to more questions in this category (Q3.7 and Q3.8), we found a gap between the actual handling of cookie notices vs their preferred way. For instance,
20\% of participants said they agreed to cookie notices in reality, when they wanted to act differently. Figure \ref{fig:How do people wish or really interact with cookie notices} shows the differences for each category.

\begin{table}[!ht]
    \centering
    \caption{Q3.3: How do you interact with cookie notices? }
    \footnotesize
    \begin{tabular}{llc} \toprule
        Category & Examples & \textit{N} \\\midrule
        Agree & Accept, Say yes, Agree & 47  \\ 
        Decline & Reject, Reject, Cancel, Disagree & 34 \\
        Ignore & Ignore, Skip it & 8 \\ 
        Edit cookie settings & Edit cookie setting/notice& 7 \\
        Get rid of it & Make it go away, Respond quickly & 7\\ 
        Use PET & Clear cookies later/regularly & 6 \\
            \bottomrule             
    \end{tabular}
    \label{tab:interact}
\end{table}

\subsection{Issues with Cookie Notices}

\begin{figure}[t]
        \input{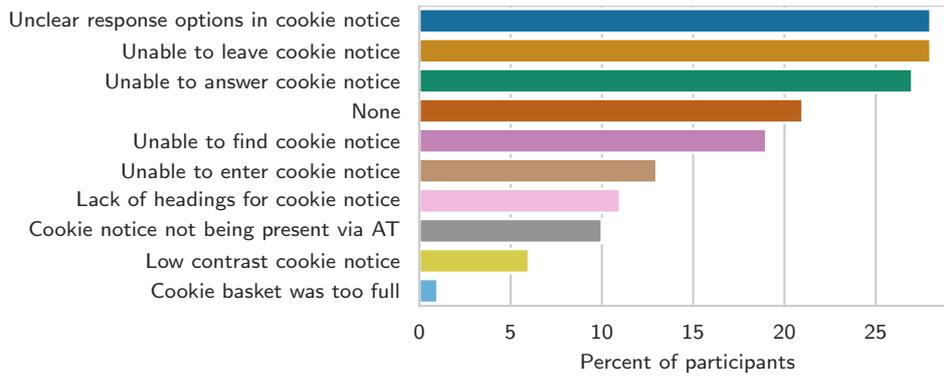}
    \caption{Q3.6: Which of the following issues have you experienced?}
    \label{fig:Which of the following issues have you experienced}
    \alt{Bar chart showing the results for Q3.6. The results are as follows: Unclear response options in cookie notice: 28, Unable to leave cookie notice: 28, Unable to answer cookie notice: 27, None: 21, Unable to find cookie notice: 19, Unable to enter cookie notice: 13, Lack of headings for cookie notice: 11, Cookie notice not being present via my AT: 10, Low contrast cookie notice: 6, Cookie basket was too full: 1.
    }
\end{figure}

For Q3.5, the majority (n=59) of participants said they had not encountered issues with cookie notices (Figure \ref{tab:describe}). 
The rest said they have experienced issues regarding cookie display or settings or described negative feelings such as frustration regarding them.
P98 said that they had experienced ``cookie notices blocking content on the page that, if not blocked, I could read and close the page without having to interact with the notice.''
P50 said ``some websites make it a bit difficult to reject all cookies, it'll open up another page where you'll have to individually select each tick box to reject.''
However, when presented with a list of possible problems in Q3.6, only 20\% said none. 
79\% of the participants said that they had experienced at least one, the most common being unclear response options in a cookie notice (n=28) and being unable to leave a cookie notice (n=28). Detailed results for this question are in Figure \ref{fig:Which of the following issues have you experienced}. 

\begin{table}[ht]
    \centering
    \caption{Q3.5: Please describe in your own words what type of issues have you experienced with cookie notices.}
    \footnotesize
    \begin{tabular}{llc} \toprule
        Category & Examples & \textit{N} \\\midrule
        None & Nothing, None, No problem&  59 \\ 
        Display problems  & Too big, Loading, Can't find, Can't read & 14 \\
        Cookie settings problems & Difficult to reject, Forced to accept & 13 \\
        Negative feelings & Too many, Tired of disabling, Annoying  & 9 \\
        Other & Cookies full, Tried to disable & 2 \\ \bottomrule
    \end{tabular}
    \label{tab:describe}
\end{table}

In a follow-up question (Q3.9), we asked what is the potential reason why participants cannot respond to a cookie notice. The responses of the participants fell into two main categories: technical issues (n=37) or malicious behaviour (n=16). Four participants explicitly mentioned issues with AT e.g., P27 said that ``Assistive technology may not be picking up a notice that has been given.'' For example, P15 said they believe it is ``because they're trying to force you to accept by pretending it's broken?''

\subsection{Suggestions}   
We asked our participants about the responsible stakeholders for accessibility and security/privacy of web services.
In this multiple-choice question, several entities came up including website developers (n=77),  policymakers (n=48), end-users (n=24), accessibility evaluation designers (n=18), and AT designers (n=15).

\begin{figure}[h]
    \input{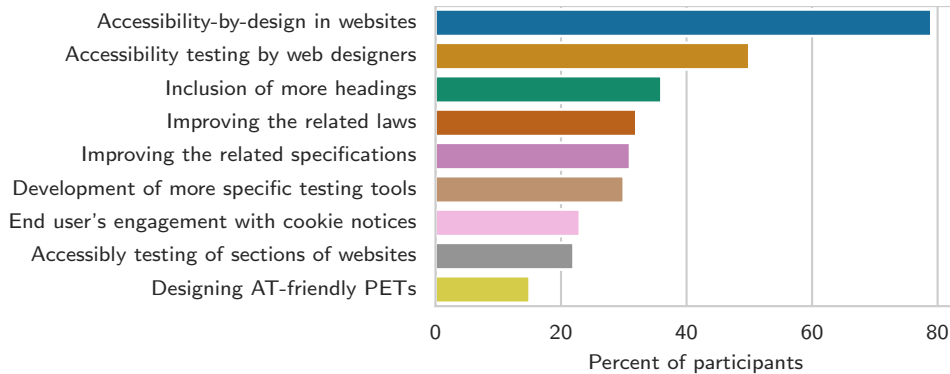}
    \caption{Q4.2: Which of these recommendations do you think would help improve your experience online? (multiple choice)}
    \label{fig:Which of the these recommendation do you think would help improve your experience online}
    \alt{Bar chart showing the results for Q4.2. The results are as follows: Accessibility-by-design in websites: 79, Accessibility testing by web designers: 50, Inclusion of more headings: 36, Improving the related laws: 32, Improving the related specifications: 31, Development of more specific testing tools: 30, End user’s engagement with cookie notices: 23, Accessibly testing of sections of the websites: 22, Designing AT-friendly PETs: 15.
    }
\end{figure}

In addition, in response to Q4.2 in which we listed a set of recommendations (based on our system studies), all participants thought that at least one of our suggestions would help to improve user experience. For example, over half of the participants believed that accessibility-by-design (n=79) in websites or more accessibility testing in websites(n=50) would help their experience. 
Figure \ref{fig:Which of the these recommendation do you think would help improve your experience online} shows the popularity of other recommendations. We discuss these in Section 7 at length. 

\section{Discussion}
In this section, we discuss our results across our system studies and user study. 

\subsection{Website Accessibility and User Privacy}
In response to RQ2-a, we found that 93.3\% of websites contained at least one accessibility error and 77.8\% contained at least one contrast error. This means that most websites tested are not compliant with the WCAG success criteria and, therefore, could be inaccessible, difficult for people with VI to access, or cause access issues.

The most common error during our testing of cookie notices was low-contrast buttons or links. The WCAG criteria 1.4.3 and 1.4.6 give guidance for contrast, the minimum guidance is a contrast ratio of at least 4.5 to 1 with enhanced guidance of a contrast ratio of at least 7 to 1 \cite{w3HowMeetWCAG2019}. For the websites that contained a contrast error, this means that they did not meet the minimum guidance and, therefore, could make text difficult to read for people with VI. Alongside this, we found several websites that had no errors in their cookie notices but contained errors elsewhere on the page. Suggesting that the overall accessibility landscape is inadequate, this aligns with previous research, e.g. \cite{hanson2013progress}.
Our results align with previous work, reporting that the majority of university library finding aids had at least one accessibility error \cite{southwellEvaluationFindingAid2013}. 

These errors with contrast could affect users who have vision loss but are not fully blind. Due to this group of people being larger than those who are fully blind, this result is concerning. Low contrast could cause users to miss important links or become confused about where to give or reject consent. For example, previous research has shown that higher contrasts between text and background colour led to faster searching \cite{ling2002effect}, as well as affecting reading speed \cite{rubinPsychophysicsReadingVI1989}. It has also been shown in a requirement survey that links can cause usability issues for users with VI \cite{yu2006novel}.

\subsection{Cookie Notice Accessibility Issues}
In response to RQ3-a, we have categorised cookie notice accessibility issues including text-only browser issues, keyboard traps, and visual presentation of cookie notices vs. screen reader output. We explain each category here. 

\textbf{Text-only Browser Issues:} WebbIE was used to manually examine the heading contained within a website and its cookie notice. We found that 58.7\% of the websites contained headings that could be useful for navigation, 15.2\% contained basic headings and 26.0\% contained no headings at all. Only 2.9\% of websites contained headings throughout their cookie notice, with 17.1\% having a heading at the start. Several cookie notices did not appear when using WebbIE, this is most likely due to WebbIE being built using the Microsoft Web Browser object which gives a program its own internal IE \cite{kingBlindPeopleWorld2004}. In June 2022, Microsoft officially ended support for IE for some OSs \cite{lifecycleFAQInternet}. It is therefore likely that web pages stopped supporting IE due to it being a legacy browser, and this caused these websites not to work with WebbIE.

\textbf{Keyboard Traps:} It was found that 77.1\% of websites that contained cookie notices were keyboard navigable when using NVDA. The most common problem found was having to intervene and use a mouse, an option that is not feasible for people with VI. There were two main times when a mouse was needed. Firstly, to get NVDA to read the cookie notice, some websites require the user to click on the cookie notice to interact with it. The other issue was escaping the cookie notice as there were websites that trapped the user in the cookie notice. This directly contradicts the WCAG success criteria 2.1.2, which is rated at level A. 
Whereas when using JAWS 82.9\% of websites that contained cookie notices were keyboard navigable. Due to how JAWS operates a higher number of privacy policies could be read, with fewer of them creating a keyboard trap. This is most likely due to how different screen readers handle CSS code differently \cite{webaimScreenReadersCSS2017}. This is an example of these privacy-enhancing technologies, required under the GDPR, directly impacting the usability of websites for people with VIs. Further work is still needed on this topic, however, this links directly to our suggestions in Section 8.3 arguing the need for accessibility and privacy requirements to act in harmony for the benefit of all users.

\textbf{Visual Presentation vs Screen Reader Audio Output:} Only 14.3\% of the cookie notices contained buttons or links whose use could be determined solely by the button or link when using NVDA. Whereas 31.4\% of the cookie notices contained buttons or links whose use could be determined solely by the button or link. This difference was due to JAWS reading out alternative text associated with buttons on some websites. An example of this is where a button might visually only say \emph{Accept all} whereas when read aloud using JAWS it says ``Accept the use of cookies and other data for the purposes described”. This change gives the user significantly more context on what the button does for them and allows them to skip the reading of the cookie notice. 
However, it could be argued that this could be done without the additional alternative text and, therefore, benefit both users without and with VI. For example, the accept button on a website simply reads \emph{Accept all cookies}.
Therefore, it was easy to ascertain the function of the button, only from this text, without the need for additional markup. 

\subsection{Reading Aloud the Cookie Notice} 
When using a screen reader, the content of the web page is spoken out loud in a linear order, which may differ from the visual order on the screen \cite{southwellEvaluationFindingAid2013}. When using WebbIE to view the web pages non-graphically, the cookie notices were often not at the start of the web page. To combat this, screen readers can also navigate using headings to jump to different sections. However, the lack of headings at the start of cookie notices makes it difficult to locate them when using this method. Screen readers can also search for content within a web page \cite{rotardTactileWebBrowsing2008}, but without a clear starting heading this becomes difficult. There were multiple websites where the cookie notice was not read aloud immediately, and the cookie notice also did not include a heading. In these examples, it would be difficult to navigate to the cookie notice, without either knowing what you are searching for or visually identifying it.

When using NVDA with Chrome, 82.9\% of cookie notices were read aloud, with 57.1\% immediately after the website title was read. There were also websites that read the cookie notice quickly after opening but not immediately; for example, elements such as navigation bars were often read aloud before cookie notices. For example, one website (ebay.co.uk) reads the title of the page, then the navigation bar, and then the cookie notice.
These were normally websites that did not display the notice at the top when using a browser graphically. In another example, graphically the cookie notice was at the bottom but was read after the heading of the website, and before the main body.
A possible reason for this is the hierarchy of the underlying HTML and CSS code.  

When using JAWS with Google Chrome, 97.1\% of cookie notices could be read aloud, with 62.9\% being read aloud immediately. The main issue when a cookie notice was not read immediately was that the user had to go through the whole page to read the cookie notice.  
As we showed in the results section, once loaded, these websites start collecting data at a large scale and even before user interaction with the cookie notice. When the cookie notice is the last item to be read to a VI user, it can easily distract the user from engaging with it leading to missing the cookie notice altogether. 

The results of our user study also confirm that cookie notice accessibility issues are indeed associated with negative feelings (RQ4). This is in line with previous research on cookie notices \cite{coopamootooFeelInvadedAnnoyed}. In response to RQ4-a, the results also highlighted that there are a range of display issues with these cookie notices, such as ``they can't be read''. This contributed to the gap we identified between the way that these users handle these cookie notices vs how they would like to handle them (Figure \ref{fig:How do people wish or really interact with cookie notices}). This also highlights the tension between the GDPR and the need for accessibility and shows a lack of work from developers to make their cookie notices accessible to users. We give solutions to the relevant parties in our recommendations (Sections 8.3 and 8.1 respectively).

\subsection{Website vs Cookie Notice Accessibility}
100\% of websites contained a title, while only 51.4\% of cookie notices contained a title, which explained what it was. This result was the same for both screen readers. This lack of titles makes it more difficult to use headings to quickly navigate to the cookie notice. It is more of a problem when the notice is not immediately read aloud and then the user has to navigate to it. The lack of a title also means the user might miss the cookie notice. 
None of the 6 websites which contained abbreviations explained them on both screen readers. This lack of explanation affects the understanding of all users and directly contradicts WCAG success criteria 3.1.4. Regardless of this being a high-level success criterion, it is important in the context of cookie notices. Adding some type of mechanism for understanding abbreviations when they are used would help all users understand what they are agreeing to. 

In response to RQ3-b, we summarise the impact of the issues we encountered on users with VI. The fact that some cookie notices were missing when using the text-only browser means that users would not be able to respond to them. This also applies to the cookie notices that were not readable using screen readers. 
Similarly, users could not consent when cookie notices are not being read immediately, not including headings, and generally being difficult to navigate.
Such a practice might require users to apply additional effort to specifically navigate to the notice. The lack of headings, structural elements, and explanatory buttons within the cookie notice means that it could take users with VI a longer time to respond to a cookie notice than other users. 
All these issues mean that these users are less protected against online tracking and cookies can be placed on their devices without any possibility for the users to know or give consent.

\section{Recommendations}
In this section, we provide a set of recommendations and best practices for different stakeholders. 

\subsection{Website Developers} 
There are several ways for websites to have the maximum compatibility with the tools and software used by people with VI. When including a cookie notice, it should be close to the top of the document's code. This will allow screen readers and other accessibility tools to quickly output this to the user. Developers could then use CSS code to change the visual location, meaning that a screen reader would always be able to read it aloud quickly. For example, developers may want visually to move it to the lower left corner (on desktop) or the lower part of the screen (on mobile) to improve the number of consent decisions for users without VI \cite{utzInformedConsentStudying2019}.
In addition, developers should always include a heading at the start of important content, whether this is a cookie notice or other important information. This allows ATs to easily and quickly navigate to this information. It also allows users to quickly understand the content of the section they are about to interact with and therefore if this information is useful to them.

To assess usability, developers should aim to use a multitude of tools. Tools such as WAVE and Lighthouse are aimed at allowing developers to easily evaluate a website. However, we showed in response to RQ2-b, they do not always highlight the problems users may face and high scores do not necessarily mean that a website is accessible. This is specifically the case when it comes to cookie notices and potentially other PETs.
Therefore, more manual tests should be undertaken to find more nuanced issues with a website. Such testing should be conducted in a comprehensive manner and by multiple VI tools, since using a combination of such tools is a common practice for VI users. The tools we suggest are WAVE to perform automated testing, followed up by using a screen reader such as NVDA since it is free and relatively easy to use.

Some of these suggestions are included in some web accessibility best practice guides, for example regarding heading use~\cite{30WebAccessibility, initiativewaiDesigningWebAccessibility2024} and accessibility testing~\cite{kristenbakerWebAccessibilityUltimate,30WebAccessibility}. However, we find that some advice goes directly against our findings and suggestions, namely cookie notices positioning~\cite{katiehempeniusBestPracticesCookie}. As stated in section 8.3, we also believe in the need for advice specifically regarding cookie notices.

\subsection{Designers of Accessibility Evaluation Tools} 
This research shows that accessibility tools and software available do not automatically assess websites for their privacy practices. The addition of the ability to test sub-sections of a website for accessibility issues would make testing elements of a website, such as a cookie notice, a simpler process. This would allow testing just to focus on such elements. In addition, this will enable subsets of development teams to test the accessibility of their work. In our testing with automated tools, it was often not clear where the errors and alerts were without further manual evaluation. However, this practice should not replace the overall accessibility testing of the website but would allow more focus to be given to some areas of the website.

Furthermore, the creation of specific accessibility tools and tests for cookie notices and other PETs would greatly improve real-world standard practices. Such tools could not only test the accessibility of the cookie notices, privacy policies, and other PETs, but also could evaluate the law compliance across platforms e.g., software, websites, and mobile apps.

\subsection{Policymakers}
To respond to RQ1-b, we have performed both accessibility and GDPR compliance analysis. 
Overall if all websites comply with WCAG, it would benefit all users, especially users with disabilities.
The question of GDPR compliance is a more complicated question in relation to users with VI. GDPR bring many benefits for the privacy of users, however, in some cases, the implementation of cookie notices has affected the overall accessibility of websites. For this reason, we make the following recommendation to policymakers.

The inclusion of specific guidelines for accessibility issues of user privacy which align with those included in GDPR and the ePD would generally improve the landscape. 
For example, guidelines on specific matters such as the length of time after loading that a cookie notice should be read aloud, what should be included in the content of the notices, and how the options should be presented to the users. WCAG version 2.2 has added limited guidance on cookie notices~\cite{gov.ukdesignsystemCookieBannern.d., w3WebContentAccessibility2023}, however, we still believe that more guidance is needed.

Standardisation bodies can create comprehensive specifications for website developers and dedicated privacy sections. For example, a W3C specification which includes all the information that developers need to comply with legal frameworks, such as GDPR or The California Consumer Privacy Act (CCPA), as well as guidelines, such as WCAG. Such specifications can be also offered by Google and Apple for app developers to improve the privacy of VI users. 

\subsection{End Users}
Generally, we believe that the onus around this issue should not be pushed onto end users, who are already a marginalised group. However, there are still additional steps users with VI could take.
End-users who are concerned regarding cookie notices can manually search for them. All of the browsers tested have a feature to search within websites.
However, such a practice might not be needed in the near future due to the ineffective nature of cookie notices on websites. Several papers have reported that cookie notices are not practical and even when the user opts out the websites still track the users. Some of these cookie notices are trackers themselves \cite{matte2020cookie}. 
As a response, Brave has recently announced that it would automatically block cookie notices altogether \cite{thebraveprivacyteamBlockingAnnoyingPrivacyharming2022}. This option could work to improve the privacy of users, along with the privacy-preserving nature of the Brave browser. Due to the browser being based on Chromium, it would likely be just as accessible as Google Chrome. However, this remains an open research problem to be tackled in the future.

In response to RQ4-b, we concluded that our participants believed that our set of recommendations could improve their online experience and privacy. Figure \Ref{fig:Which of the these recommendation do you think would help improve your experience online} displays the popularity of each item where accessibility-by-design in websites is rated top, followed by accessibility testing by web designers, the inclusion of more headings, improvement of related laws/specifications, development of more specific testing tools, end-user engagement with cookie notices, accessibility testing of sections of websites (including cookie notices), and designing AT-friendly PETs.

\section{Conclusion}
This paper investigated the interaction between ATs and cookie notices via a set of system studies of 46 top UK websites and a user study of 100 users with VI via Prolific Academic. 
We find that 22 of these websites had at least one issue with the accessibility of their cookie notice when manually tested using a screen reader. We also observed websites which did not have issues with their cookie notices when using AT but did include issues such as low contrast when viewing them graphically. 
These practices often created accessibility issues when trying to read and respond to cookie notices.
The results of our user study revealed that users with VI overall have a negative view of cookie notices. We also find that all users believe that at least one of our recommendations would help improve their experience online.

In future work, we would like to conduct cross-platform studies looking at mobile web browsers, mobile apps, and desktop web browsers and their interaction with AT. We would also like to automate our methodology and run large-scale system studies. 
We also would like to focus on the creation and adaptation of dedicated accessibility testing tools for privacy matters and compliance with the law.

\section*{Acknowledgements}

The second and third authors are part of research supported by the UK Research and Innovation (UKRI), through the Strategic Priority Fund as part of the Protecting Citizens Online programme (Grant: AGENCY: Assuring Citizen Agency in a World with Complex Online Harms, EP/W032481/2).

\bibliographystyle{ACM-Reference-Format}
\bibliography{new_bib}

\appendix
\newcounter{step}
\setcounter{step}{1} 
\section{Website Analysis Template} \label{website_analysis_template}
For each website in our list, we followed these steps for our analysis:
\begin{itemize}
    \footnotesize
    \item \textbf{Part I: Cookie Notice (baseline)}
    \begin{itemize}
        \item[-] Step \thestep \refstepcounter{step}: \label{cookie_notice} Open Google Chrome incognito window and visit the homepage of the website.
        \item[-] Step \thestep \refstepcounter{step}: Observe if there is a notice (cookie consent, privacy settings, banner, etc.). 
        \begin{itemize}
            \item[-] No: Write it in the file.
            \item[-] Yes: Observe the location and user control options e.g. OK, Accept, Yes, Reject, No, More Options, Settings, Links to privacy-related pages, etc. Write your observations in the file.
        \end{itemize}
        \item[-] Step \thestep \refstepcounter{step}: Close the Google Chrome incognito window.
        \item[-] Step \thestep \refstepcounter{step}: Open Brave private window and visit the homepage of the website.
        \item[-] Step \thestep \refstepcounter{step}: Repeat Step \ref{cookie_notice} (cookie notice).
        \item[-] Step \thestep \refstepcounter{step}: Record the number of items blocked by the Brave Shields in the file.
        \item[-] Step \thestep \refstepcounter{step}: Close the Brave private window.
    \end{itemize}
    \item \textbf{Part II: Automated Accessibility Testing Tools}
    \begin{itemize}
        \item[-] Step \thestep \refstepcounter{step}: Open a new Google Chrome incognito window and visit the website's homepage.
        \item[-] Step \thestep \refstepcounter{step}: Click on WAVE extension to run the test.
        \item[-] Step \thestep \refstepcounter{step}: Record the number of each of the categories in the file.
        \item[-] Step \thestep \refstepcounter{step}: Close the Google Chrome incognito window.
        \item[-] Step \thestep \refstepcounter{step}: Open a new Google Chrome incognito window and visit the website's homepage.
        \item[-] Step \thestep \refstepcounter{step}: Open developer tools and navigate to the Lighthouse tab.
        \item[-] Step \thestep \refstepcounter{step}: Select Navigation Mode, A desktop device and the Accessibility Categories.
        \item[-] Step \thestep \refstepcounter{step}: Analyse the page.
        \item[-] Step \thestep \refstepcounter{step}: Record the overall accessibility score and the number of each item shown.
        \item[-] Step \thestep \refstepcounter{step}: Close the Google Chrome incognito window.
    \end{itemize}
    \item \textbf{Part III: Manual Testing Via Text-only Browser}
    \begin{itemize}
        \item[-] Step \thestep \refstepcounter{step}: Open WebbIE Browser and visit the homepage of the website.
        \item[-] Step \thestep \refstepcounter{step}: Record the number of headings for the website overall.
        \item[-] Step \thestep \refstepcounter{step}: Record the presence of a notice, and if so the presence of headings.
        \item[-] Step \thestep \refstepcounter{step}: Close WebbIE.
    \end{itemize}
    \item \textbf{Part IV: Manual Testing via Screen Readers}
    \begin{itemize}
        \item[-] Step \thestep \refstepcounter{step} \label{reader_start}: Open NVDA screen reader.
        \item[-] Step \thestep \refstepcounter{step}: Open a new Google Chrome incognito window and visit the website's homepage.\
        \item[-] Step \thestep \refstepcounter{step}: Allow screen reader to read website.
        \item[-] Step \thestep \refstepcounter{step}: Interact with the website using keyboard.
        \item[-] Step \thestep \refstepcounter{step}:\label{reader_finish} Record pass/fail for categories in 3.2.3.
        \item[-] Step \thestep \refstepcounter{step}: Close the Google Chrome incognito window and screen reader. 
        \item[-] Step \thestep \refstepcounter{step}: Open JAWS screen reader.
        \item[-] Step \thestep \refstepcounter{step}: Repeat Steps \ref{reader_start}-\ref{reader_finish}.
    \end{itemize}
\end{itemize}

\noindent Note that we performed two rounds of testing (with identical results). We uninstalled and reinstalled WebbIE for each round since it does not support cookie management. 
\section{User Study Template} \label{questionnaire}
\footnotesize
    \subsection{Screening validation}
    \begin{itemize}
        \item Do you live in the United Kingdom? (One answer is possible)

        \begin{checkboxes}
            \item Yes
            \item No
        \end{checkboxes}

        \item Do you use assistive technology? (One answer is possible)
        \begin{checkboxes}
            \item Yes
            \item No
        \end{checkboxes}
    \end{itemize}
    \subsection{Section 1: Internet and assistive technology}

    This section is about your internet usage and any assistive technology you may utilise while browsing the web.
    \begin{itemize}
        \item 1.1: How do you describe your visual impairment? (Text input is possible)
        
        \item 1.2: Which devices do you use to access the internet? Please specify if other (Several answers are possible)
        
        \begin{checkboxes}
            \item Personal Computer (Desktop or Laptop) Mobile Phone
            \item Tablet Computer
            \item Smart home devices
            \item Wearable devices Internet-enabled TVs Gaming consoles Public computers
            \item Other: \underline{\hspace{5cm}}
        \end{checkboxes}
        
        \item 1.3: Which device are you using to complete this questionnaire? Please specify if other (One answer possible)
        \begin{checkboxes}
            \item Personal Computer (Desktop or Laptop)
            \item Mobile Phone
            \item Tablet Computer
            \item Smart home devices
            \item Wearable devices 
            \item Internet-enabled TVs 
            \item Gaming consoles 
            \item Public computers 
            \item Other: \underline{\hspace{5cm}}
        \end{checkboxes}
        
        \item 1.4: In an average week, how many hours do you use the internet? (One answer is possible)
        \begin{checkboxes}
            \item Less than 1 hour 1-5 hours
            \item 6-10 hours
            \item 11-15 hours
            \item 16-20 hours
            \item 21-15 hours
            \item 26-30 hours
            \item More than 30 hours
        \end{checkboxes}
        
        \item 1.5: What assistive technology do you use when browsing the web? Please specify if other. (Several answers are possible)
        \begin{checkboxes}
            \item Screen reader (Jaws, NVDA, Voice over or other)
            \item Braille Display
            \item Text Only Browser (WebbIE or other) 
            \item Magnification software
            \item Assistive browser extension 
            \item Alternative input devices 
            \item None
            
        \end{checkboxes}

        \item 1.6: Which of the following screen readers do you use? Please specify if other. (Several answers are possible)
        \begin{checkboxes}
            \item JAWS
            \item NVDA 
            \item VoiceOver 
            \item Natural Reader 
            \item Read\&Write 
            \item Narrator 
            \item Talkback 
            \item ChromeVox 
            \item Orca
            \item I don't use a screen reader
            \item Other: \underline{\hspace{5cm}}
        \end{checkboxes}    

        \item 1.7: How would you describe your level of expertise with a screen reader? (Text input is possible)

        \item 1.8: Do you use plug-ins with a screen reader? (One answer is possible)

        A plugin or add-on adds a specific feature or functionality to a screen reader. Allowing users to customise their software and add the features they need.
        \begin{checkboxes}
            \item Yes
            \item No
        \end{checkboxes}

        \item 1.9: Which plugins do you use? (Text input is possible)
    \end{itemize}

    \subsection{Section 2: Privacy enhancing technology usage}
        This section is about the measures you take to protect your privacy and security while browsing the internet and Privacy Enhancing Technologies (PETs). PETs are tools that can help protect your privacy online by limiting the collection, use, and dissemination of your personal information.
        \begin{itemize}
            \item 2.1: Which of the following privacy-enhancing technologies do you use? Please specify if other. (Several answers are possible)
                \begin{checkboxes}
                    \item Browsers do not track
                    \item Virtual Private Network (VPN) 
                    \item Private browsing
                    \item Password manager
                    \item Manual cookie opt-out
                    \item Privacy-focused web browsers 
                    \item Encrypted messaging apps
                    \item Ad blockers
                    \item File encryption tools
                    \item None
                    \item Other: \underline{\hspace{5cm}}
                \end{checkboxes}

            \item 2.2: How did you learn about privacy-enhancing technology? Please specify if other. (Several answers are possible)
            \begin{checkboxes}
                \item Friend/social contact recommendation 
                \item Work/school recommendation 
                \item Privacy/cookie policy of a website 
                \item News
                \item I don't know
                \item I don't use privacy-enhancing technologies
                \item Other: \underline{\hspace{5cm}}
            \end{checkboxes}
        \end{itemize}

        \subsection{Section 3: Cookie notice}

        Cookie notices appear when a user lands on a website and informs them that the website is using cookies (a data file) and other trackers that process personal data and that the user must choose whether they want their personal data processed.

        \begin{itemize}
            \item 3.1: In your own words what is a cookie notice and what are they supposed to do? (Text input is possible)
            \item 3.2: How do you feel about cookie notices? (Text input is possible)
            \item 3.3: How do you interact with cookie notices? (Text input is possible)
            \item 3.4: Have you encountered any issues with cookie notices? (One answer is possible)
            \begin{checkboxes}
                \item Yes
                \item No
            \end{checkboxes}
            \item 3.5: Please describe in your own words what type of issues have you experienced with cookie notices. (Text input is possible)
            \item 3.6: Which of the following issues have you experienced? Please specify if other (Several answers are possible)
            \begin{checkboxes}
                \item Cookie notice not being present via my assistive technology 
                \item Unable to find cookie notice
                \item Unable to answer cookie notice
                \item Low contrast cookie notice
                \item Unclear response options in the cookie notice 
                \item Lack of headings for cookie notice 
                \item Unable to leave cookie notice
                \item Unable to enter cookie notice
                \item Other: \underline{\hspace{5cm}}
            \end{checkboxes}
            \item 3.7: How would you wish to handle cookie notices? Please specify if other. (One answer is possible)
            \begin{checkboxes}
                \item Agree
                \item Disagree
                \item Editing the settings 
                \item Ignore cookie notice
                \item Other: \underline{\hspace{5cm}}
            \end{checkboxes}
            \item 3.8: How do you actually respond to cookie notices? Please specify if other. (One answer is possible)
            \begin{checkboxes}
                \item Agree
                \item Disagree
                \item Editing the settings
                \item Ignore cookie notice
                \item Unable to respond to cookie notice
                \item Other: \underline{\hspace{5cm}}
            \end{checkboxes}
            \item 3.9: If you are unable to respond to cookie notices, what do you think is the reason? (Text input is possible)
        \end{itemize}

    \subsection{Section 4: Suggestions}
    \begin{itemize}
        \item 4.1: Who do you think should be responsible for ensuring the secure accessibility of the Internet? Please specify if other. (Several answers are possible)
        \begin{checkboxes}
            \item Website Developers
            \item Designers of Accessibility Evaluation Tools 
            \item Policymakers
            \item Users of the internet
            \item Designers of Assistive Technologies
            \item Other: \underline{\hspace{5cm}}
        \end{checkboxes}

        \item 4.2: Which of these recommendations do you think would help improve your experience online? Please specify if other (Several answers are possible)
        \begin{checkboxes}
            \item Website designers should design websites with accessibility in mind
            \item Website designers should include more headings for useful information
            \item Website designers should complete more accessibility testing
            \item Evaluation tools should allow for testing of sections of a web page
            \item Specific testing tools for parts of a webpage (i.e. cookie notices or other elements)
            \item The inclusion of laws specifying privacy and accessibility
            \item Specifications of achieving privacy and accessibility
            \item End users can search for cookie notices
            \item End users can specific privacy-protecting tools such as the Brave Internet browser
            \item Other: \underline{\hspace{5cm}}
        \end{checkboxes}    
    \end{itemize}

    \subsection{Section 5: Demographic and background questions}

    \begin{itemize}
        \item 5.1: What is your age? (One answer is possible)

        \begin{checkboxes}
            \item 18 to 24
            \item 25 to 34
            \item 35 to 44
            \item 45 to 54
            \item 55 to 64
            \item 65 or over
            \item Prefer not to say
        \end{checkboxes}

        \item 5.2: What is your gender? (One answer is possible)
        \begin{checkboxes}
            \item Female
            \item Male
            \item Non-binary
            \item Prefer not to say
            \item Other
        \end{checkboxes}

        \item 5.3: What is your highest level of education? (One answer is possible)
        \begin{checkboxes}
            \item Secondary education
            \item Post-secondary education
            \item Undergraduate Education
            \item Graduate Education
            \item Prefer not to say
        \end{checkboxes}

        \item 5.4: What is your occupation? Please specify if other. (One answer is possible)
        \begin{checkboxes}
            \item Healthcare and social assistance
            \item Education and training
            \item Sales and retail
            \item Administrative and support
            \item Manufacturing and production
            \item Information technology
            \item Business and finance
            \item Transportation and logistics
            \item Construction and trades
            \item Arts, entertainment, and media
            \item Prefer not to say
            \item Other
        \end{checkboxes}

        \item 5.5: What services do you use online? Please specify if other. (Multiple answers are possible)
        \begin{checkboxes}
            \item Email
            \item Social media
            \item Online shopping and e-commerce
            \item Video Streaming
            \item Music/audio streaming
            \item Online payment/banking
            \item File sharing and cloud storage
            \item Online travel booking
            \item Online education and e-learning
            \item Online commutation and collaboration services
            \item Other
        \end{checkboxes}
    \end{itemize}
\begin{table*}[htbp]
\section{Detailed results of automated accessibility tools}
    \centering
\footnotesize\settowidth\rotheadsize{Lighthouse}
    \renewcommand\rotheadgape{}
    \caption{Detailed results of WAVE 5.0 and Google Lighthouse} \label{tab7} 
    \begin{tabular}{lrrrrrrrr} \toprule
    Website & \rothead{Privacy Policy} & \rothead{Errors} & \rothead{Contrast Errors} & \rothead{Alerts} & \rothead{Features} & \rothead{Structural Elements} & \rothead{ARIA} & \rothead{Lighthouse Score} \\ \midrule
    google.com  & \cmark  & 1     & 10    & 4     & 5     & 7     & 350   & 97 \\
    youtube.com & \cmark  & 26    & 1     & 70    & 92    & 66    & 676   & 89 \\    
    yahoo.com & \cmark  & 2     & 8     & 6     & 2     & 4     & 4     & 86 \\
    facebook.com & \cmark  & 4     & 46    & 12    & 1     & 9     & 9     & 93 \\
    bbc.co.uk & \cmark  & 0     & 0     & 134   & 23    & 119   & 371   & 100 \\
    amazon.co.uk & \cmark  & 6     & 2     & 139   & 194   & 55    & 381   & 95 \\
    reddit.com & \cmark  & 9     & 139   & 657   & 104   & 36    & 433   & 77 \\
    wikipedia.org & \xmark & 3     & 0     & 97    & 123   & 70    & 24    & 100 \\
    live.com & \xmark & 0     & 5     & 8     & 11    & 38    & 77    & 98 \\
    instagram.com & \cmark  & 0     & 25    & 3     & 10    & 10    & 37    & 95 \\
    twitter.com & \cmark  & 1     & 19    & 3     & 1     & 4     & 55    & 88 \\
    ebay.co.uk & \xmark & 0     & 0     & 0     & 0     & 0     & 0     & 93 \\
    dailymail.co.uk & \cmark  & 75    & 31    & 890   & 628   & 497   & 61    & 74 \\
    bing.com & \cmark  & 12    & 23    & 38    & 26    & 40    & 244   & 94 \\
    gov.uk & \cmark  & 0     & 1     & 5     & 16    & 68    & 38    & 100 \\
    netflix.com & \cmark  & 2     & 22    & 9     & 22    & 36    & 68    & 88 \\
    theguardian.com & \xmark & 17    & 59    & 131   & 102   & 290   & 624   & 79 \\
    pornhub.com & \xmark & 65    & 1     & 197   & 110   & 67    & 119   & 96 \\
    office.com & \cmark  & 4     & 0     & 3     & 25    & 59    & 199   & 94 \\
    fandom.com & \cmark  & 12    & 33    & 14    & 28    & 28    & 2     & 86 \\
    xvideos.com & \cmark  & 104   & 64    & 158   & 2     & 22    & 1     & 88 \\
    paypal.com & \cmark  & 12    & 33    & 14    & 28    & 28    & 2     & 100 \\
    microsoft.com & \cmark  & 2     & 0     & 3     & 24    & 47    & 183   & 100 \\
    linkedin.com & \cmark  & 27    & 0     & 0     & 9     & 32    & 174   & 100 \\
    xhamster.com & \xmark & 8     & 118   & 57    & 58    & 25    & 2     & 81 \\
    imdb.com & \xmark & 8     & 0     & 42    & 98    & 37    & 1552  & 88 \\
    duckduckgo.com & \xmark & 23    & 5     & 13    & 6     & 38    & 35    & 96 \\
    amazon.com & \xmark & 4     & 2     & 178   & 216   & 45    & 317   & 98 \\
    zoom.us & \cmark  & 7     & 3     & 58    & 64    & 93    & 889   & 80 \\
    twitch.tv & \cmark  & 71    & 0     & 150   & 130   & 48    & 300   & 86 \\
    amazonaws.com & \cmark  & 4     & 14    & 140   & 228   & 377   & 87    & 86 \\
    tiktok.com & \cmark  & 52    & 14    & 54    & 9     & 59    & 0     & 63 \\
    whatsapp.com & \cmark  & 5     & 14    & 10    & 4     & 142   & 5     & 85 \\
    doubleclick.net & \cmark  & 2     & 4     & 51    & 11    & 78    & 141   & 100 \\
    spankbang.com & \xmark & 16    & 160   & 500   & 190   & 36    & 0     & 72 \\
    sky.com & \cmark  & 3     & 16    & 28    & 22    & 41    & 105   & 90 \\
    apple.com & \xmark & 11    & 0     & 15    & 47    & 62    & 217   & 92 \\
    rightmove.co.uk & \cmark  & 3     & 7     & 68    & 12    & 48    & 55    & 87 \\
    booking.com & \cmark  & 30    & 3     & 37    & 40    & 85    & 574   & 92 \\
    etsy.com & \cmark  & 7     & 1     & 31    & 42    & 175   & 1682  & 73 \\
    indeed.com & \cmark  & 2     & 0     & 8     & 20    & 25    & 134   & 90 \\
    msn.com & \cmark  & 26    & 0     & 213   & 359   & 311   & 44    & 79 \\
    github.com & \xmark & 2     & 35    & 18    & 114   & 74    & 179   & 86 \\
    adobe.com & \cmark  & 2     & 1     & 12    & 111   & 120   & 139   & 96 \\
    chaturbate.com & \xmark & 19    & 113   & 494   & 102   & 210   & 5     & 84 \\
    xnxx.com & \cmark  & 165   & 0 & 808   & 3     & 20    & 0     & 97 \\ \bottomrule
    \end{tabular}
\end{table*}

\begin{table*}[htbp]
\section{Detailed results of screen reader tests}
    \begin{adjustwidth}{-.5in}{-.5in}  

\centering
\scriptsize
    \caption{NVDA and JAWS results}       
    \label{tab11}
    \settowidth\rotheadsize{Abbreviations}
    \renewcommand\rotheadgape{}
    \begin{tabular}{l m{0.1cm}m{0.1cm}m{0.4cm}m{0.4cm}m{0.4cm}m{0.1cm}m{0.4cm}m{0.6cm}m{0.2cm} |m{0.1cm}m{0.1cm}m{0.4cm}m{0.4cm}m{0.4cm}m{0.1cm}m{0.4cm}m{0.6cm}m{0.1cm}}\toprule
          & \multicolumn{9}{c}{NVDA} & \multicolumn{9}{c}{JAWS} \\ \cmidrule(r){2-10} \cmidrule(r){11-19}
    Website & \rothead{\scriptsize Readable} &\rothead{\scriptsize Immediately} & \rothead{\scriptsize Keyboard Navigable} & \rothead{\scriptsize Link or button Purpose} & \rothead{\scriptsize Abbreviations are explained} & \rothead{\scriptsize Page Titled} & \rothead{\scriptsize Cookie Notice Titled} & \rothead{\scriptsize Headings useful for navigation} & \rothead{\scriptsize No Timing} & \rothead{\scriptsize Readable} &\rothead{\scriptsize Immediately} & \rothead{\scriptsize Keyboard Navigable} & \rothead{\scriptsize Link or button Purpose} & \rothead{\scriptsize Abbreviations are explained} & \rothead{\scriptsize Page Titled} & \rothead{\scriptsize Cookie Notice Titled} & \rothead{\scriptsize Headings useful for navigation} & \rothead{\scriptsize No Timing}  \\ \midrule
    google.com  & \cmark  & \cmark  & \cmark  & \xmark & -   & \cmark  & \xmark & \xmark & \cmark  & \cmark  & \cmark  & \cmark  & \xmark & -   & \cmark  & \xmark & \xmark & \cmark \\
    youtube.com & \cmark  & \cmark  & \cmark  & \xmark & -   & \cmark  & \xmark & \xmark & \cmark  & \cmark  & \cmark  & \cmark  & \cmark  & -   & \cmark  & \xmark & \xmark & \cmark \\
    yahoo.com & \cmark  & \cmark  & \cmark  & \xmark & \xmark & \cmark  & \cmark  & \xmark & \cmark  & \cmark  & \cmark  & \cmark  & \xmark & \xmark & \cmark  & \cmark  & \xmark & \cmark \\
    facebook.com & \cmark  & \xmark & \cmark  & \xmark & \xmark & \cmark  & \cmark  & \xmark & \cmark  & \cmark  & \xmark & \cmark  & \cmark  & \xmark & \cmark  & \cmark  & \cmark  & \cmark \\
    bbc.co.uk & \cmark  & \cmark  & \cmark  & \xmark & -   & \cmark  & \cmark  & \xmark & \cmark  & \cmark  & \cmark  & \cmark  & \xmark & -   & \cmark  & \cmark  & \xmark & \cmark \\
    amazon.co.uk & \cmark  & \cmark  & \cmark  & \xmark & -   & \cmark  & \cmark  & \xmark & \cmark  & \cmark  & \xmark & \cmark  & \cmark  & -   & \cmark  & \cmark  & \xmark & \cmark \\
    reddit.com & \xmark & \xmark & \xmark & \xmark & -   & \cmark  & \xmark & \xmark & \cmark  & \xmark & \xmark & \xmark & \xmark & -   & \cmark  & \xmark & \xmark & \cmark \\
    wikipedia.org & - & - & - & - & -   & \cmark  & - & - & - & - & - & - & - & -   & \cmark  & - & - & - \\
    live.com & - & - & - & - & -   & \cmark  & - & - & - & - & - & - & - & -   & \cmark  & - & - & - \\
    instagram.com & \xmark & \xmark & \cmark  & \xmark & -   & \cmark  & \cmark  & \xmark & \cmark  & \cmark  & \xmark & \cmark  & \cmark  & -   & \cmark  & \cmark  & \cmark  & \cmark \\
    twitter.com & \cmark  & \cmark  & \cmark  & \cmark  & -   & \cmark  & \cmark  & \xmark & \cmark  & \cmark  & \cmark  & \cmark  & \cmark  & -   & \cmark  & \cmark  & \xmark & \cmark \\
    ebay.co.uk & \cmark  & \xmark & \cmark  & \xmark & \xmark & \cmark  & \xmark & \xmark & \cmark  & \cmark  & \xmark & \cmark  & \cmark  & \xmark & \cmark  & \xmark & \xmark & \cmark \\
    dailymail.co.uk & \xmark & \xmark & \xmark & \xmark & \xmark & \cmark  & \xmark & \xmark & \cmark  & \cmark  & \xmark & \xmark & \xmark & \xmark & \cmark  & \xmark & \xmark & \cmark \\
    bing.com & \cmark  & \xmark & \cmark  & \xmark & -   & \cmark  & \xmark & \xmark & \cmark  & \cmark  & \xmark & \cmark  & \xmark & -   & \cmark  & \xmark & \xmark & \cmark \\
    gov.uk & \cmark  & \cmark  & \cmark  & \cmark  & -   & \cmark  & \cmark  & \xmark & \cmark  & \cmark  & \cmark  & \cmark  & \cmark  & -   & \cmark  & \cmark  & \xmark & \cmark \\
    netflix.com & \cmark  & \cmark  & \cmark  & \xmark & -   & \cmark  & \xmark & \xmark & \cmark  & \cmark  & \cmark  & \cmark  & \xmark & -   & \cmark  & \xmark & \xmark & \cmark \\
    theguardian.com & \cmark  & \xmark & \cmark  & \xmark & \xmark & \cmark  & \xmark & \xmark & \cmark  & \cmark  & \xmark & \cmark  & \xmark & \xmark & \cmark  & \xmark & \xmark & \cmark \\
    pornhub.com & - & - & - & - & -   & \cmark  & - & - & - & - & - & - & - & -   & \cmark  & - & - & - \\
    office.com & \cmark  & \cmark  & \cmark  & \xmark & -   & \cmark  & \xmark & \xmark & \cmark  & \cmark  & \cmark  & \cmark  & \xmark & -   & \cmark  & \xmark & \xmark & \cmark \\
    fandom.com & \cmark  & \xmark & \xmark & \xmark & \xmark & \cmark  & \cmark  & \xmark & \cmark  & \cmark  & \xmark & \xmark & \xmark & \xmark & \cmark  & \cmark  & \xmark & \cmark \\
    xvideos.com & \xmark & \xmark & \xmark & \xmark & -   & \cmark  & \xmark & \xmark & \cmark  & \cmark  & \cmark  & \cmark  & \xmark & -   & \cmark  & \xmark & \xmark & \cmark \\
    paypal.com & \cmark  & \xmark & \xmark & \xmark & \xmark & \cmark  & \cmark  & \xmark & \cmark  & \cmark  & \xmark & \xmark & \xmark & \xmark & \cmark  & \cmark  & \xmark & \cmark \\
    microsoft.com & \cmark  & \cmark  & \cmark  & \xmark & -   & \cmark  & \xmark & \xmark & \cmark  & \cmark  & \cmark  & \cmark  & \xmark & -   & \cmark  & \xmark & \xmark & \cmark \\
    linkedin.com & \cmark  & \cmark  & \cmark  & \xmark & -   & \cmark  & \cmark  & \xmark & \cmark  & \cmark  & \cmark  & \cmark  & \xmark & -   & \cmark  & \cmark  & \xmark & \cmark \\
    xhamster.com & - & - & - & - & -   & \cmark  & - & - & - & - & - & - & - & -   & \cmark  & - & - & - \\
    imdb.com & - & - & - & - & -   & \cmark  & - & - & - & - & - & - & - & -   & \cmark  & - & - & - \\
    duckduckgo.com & - & - & - & - & -   & \cmark  & - & - & - & - & - & - & - & -   & \cmark  & - & - & - \\
    amazon.com & - & - & - & - & -   & \cmark  & - & - & - & - & - & - & - & -   & \cmark  & - & - & - \\
    zoom.us & \cmark  & \cmark  & \cmark  & \xmark & -   & \cmark  & \xmark & \xmark & \cmark  & \cmark  & \cmark  & \cmark  & \cmark  & -   & \cmark  & \xmark & \xmark & \cmark \\
    twitch.tv & \cmark  & \xmark & \cmark  & \cmark  & -   & \cmark  & \cmark  & \xmark & \cmark  & \cmark  & \cmark  & \cmark  & \cmark  & -   & \cmark  & \cmark  & \xmark & \cmark \\
    amazonaws.com & \cmark  & \cmark  & \cmark  & \xmark & -   & \cmark  & \cmark  & \xmark & \cmark  & \cmark  & \cmark  & \cmark  & \cmark  & -   & \cmark  & \cmark  & \xmark & \cmark \\
    tiktok.com & \cmark  & \xmark & \xmark & \xmark & -   & \cmark  & \cmark  & \xmark & \cmark  & \cmark  & \xmark & \xmark & \xmark & -   & \cmark  & \cmark  & \xmark & \cmark \\
    whatsapp.com & \cmark  & \cmark  & \cmark  & \xmark & -   & \cmark  & \cmark  & \xmark & \cmark  & \cmark  & \cmark  & \cmark  & \xmark & -   & \cmark  & \cmark  & \xmark & \cmark \\
    doubleclick.net & \cmark  & \xmark & \cmark  & \xmark & -   & \cmark  & \xmark & \xmark & \cmark  & \cmark  & \cmark  & \xmark & \xmark & -   & \cmark  & \xmark & \xmark & \cmark \\
    spankbang.com & - & - & - & - & -   & \cmark  & - & - & - & - & - & - & - & -   & \cmark  & - & - & - \\
    sky.com & \cmark  & \cmark  & \cmark  & \xmark & -   & \cmark  & \cmark  & \xmark & \cmark  & \cmark  & \cmark  & \cmark  & \xmark & -   & \cmark  & \cmark  & \xmark & \cmark \\
    apple.com & - & - & - & - & -   & \cmark  & - & - & - & - & - & - & - & -   & \cmark  & - & - & - \\
    rightmove.co.uk & \cmark  & \cmark  & \cmark  & \cmark  & -   & \cmark  & \cmark  & \xmark & \cmark  & \cmark  & \cmark  & \cmark  & \xmark & -   & \cmark  & \cmark  & \xmark & \cmark \\
    booking.com & \cmark  & \cmark  & \cmark  & \xmark & -   & \cmark  & \cmark  & \xmark & \cmark  & \cmark  & \cmark  & \cmark  & \xmark & -   & \cmark  & \cmark  & \xmark & \cmark \\
    etsy.com & \cmark  & \cmark  & \cmark  & - & -   & \cmark  & - & - & \cmark  & \cmark  & \cmark  & \cmark  & \xmark & -   & \cmark  & - & \xmark & \cmark \\
    indeed.com & \cmark  & \cmark  & \cmark  & \cmark  & -   & \cmark  & \xmark & \xmark & \cmark  & \cmark  & \xmark & \cmark  & \cmark  & -   & \cmark  & \xmark & \xmark & \cmark \\
    msn.com & \cmark  & \cmark  & \cmark  & \xmark & -   & \cmark  & \cmark  & \xmark & \cmark  & \cmark  & \cmark  & \cmark  & \xmark & -   & \cmark  & \cmark  & \xmark & \cmark \\
    github.com & - & - & - & - & -   & \cmark  & - & - & - & - & - & - & - & -   & \cmark  & - & - & - \\
    adobe.com & \xmark & \xmark & \xmark & \xmark & -   & \cmark  & \xmark & \xmark & \cmark  & \cmark  & \xmark & \cmark  & \xmark & -   & \cmark  & \xmark & \xmark & \cmark \\
    chaturbate.com & - & - & - & - & -   & \cmark  & - & - & - & - & - & - & - & -   & \cmark  & - & - & - \\
    xnxx.com & \xmark & \xmark & \xmark & \xmark & -   & \cmark  & \xmark & \xmark & \cmark  & \cmark  & \cmark  & \cmark  & \xmark & -   & \cmark  & \xmark & \xmark & \cmark \\ \bottomrule
    \end{tabular}%
    \end{adjustwidth}
\end{table*}%

\clearpage
\section{Examples of websites, cookie notices and their outcome via screen readers}
\begin{figure}[!htbp]
    \centering
    \scriptsize
    \begin{tabular}{|lp{7cm}|}
     \multicolumn{2}{c} {\includegraphics[scale=.7]{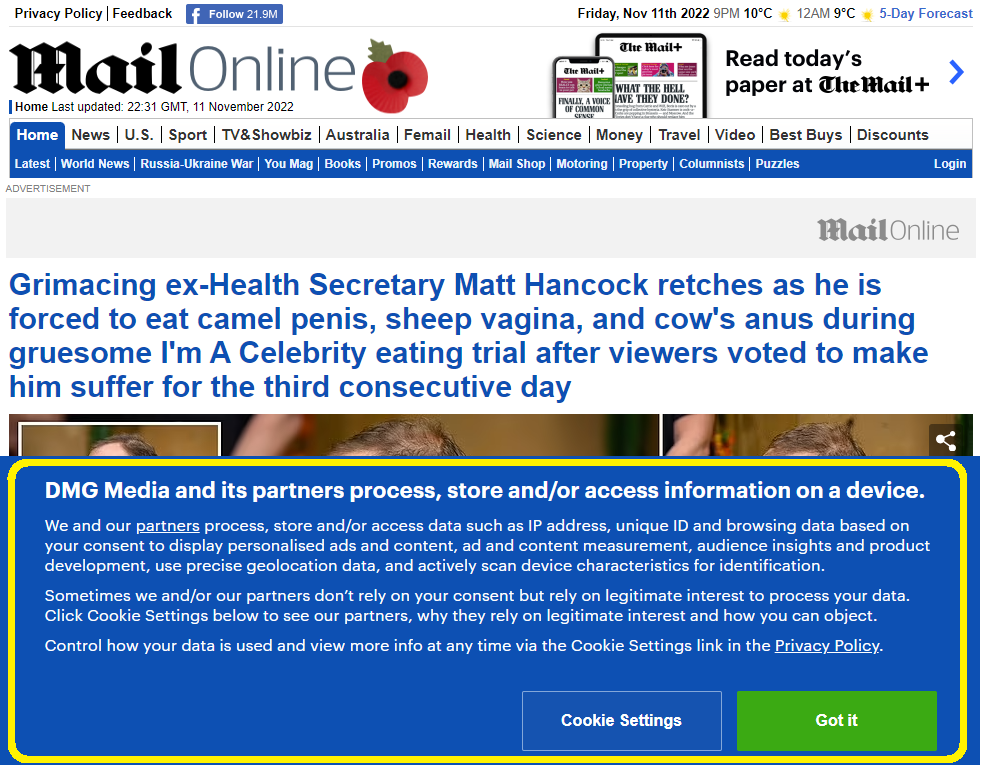}\alt{Visual cookie notice of dailymail.co.uk, offering two options: ``Cookie settings'' and ``Got it''. The cookie notice covers a large section of the lower portion of the web page.}}\\
        \multicolumn{2}{l}{(Enter webpage URL and press Enter Key)} \\\hline
        NVDA: & dailymail.co.uk selected \\
        NVDA: & UK home daily mail online \\
        NVDA: & Link, graphic online news, sport, celebrity, science and health stories\\
        NVDA: & List with 14 items \\
        NVDA: & Link home \\
        NVDA: & Link, news \\
        NVDA: & Link u.s\\
        NVDA: & [Advertisement read], [The same advertisement read]\\
        NVDA: & [Advertisement read], [The same advertisement read]\\
        NVDA: & [Advertisement read], [The same advertisement read]\\
        NVDA: & Discover the best black Friday deals, discover the best black Friday deals\\
        \multicolumn{2}{|l|}{(Down arrow key pressed)} \\
        NVDA: & Link home \\\hline
        \multicolumn{2}{l}{(Proceeds to read the navigation bar)} \\
        \multicolumn{2}{l}{(Starts to read news items on homepage)} \\
    \end{tabular}
    \caption{Top: Graphical representation of dailymail.co.uk with the highlighted cookie notice at the bottom of the page. Bottom: Example transcript via a screen reader.} \label{fig:dailymail}
\end{figure}  

\begin{figure}[!htbp]
    \scriptsize
    \centering  
    \includegraphics[scale=.15]{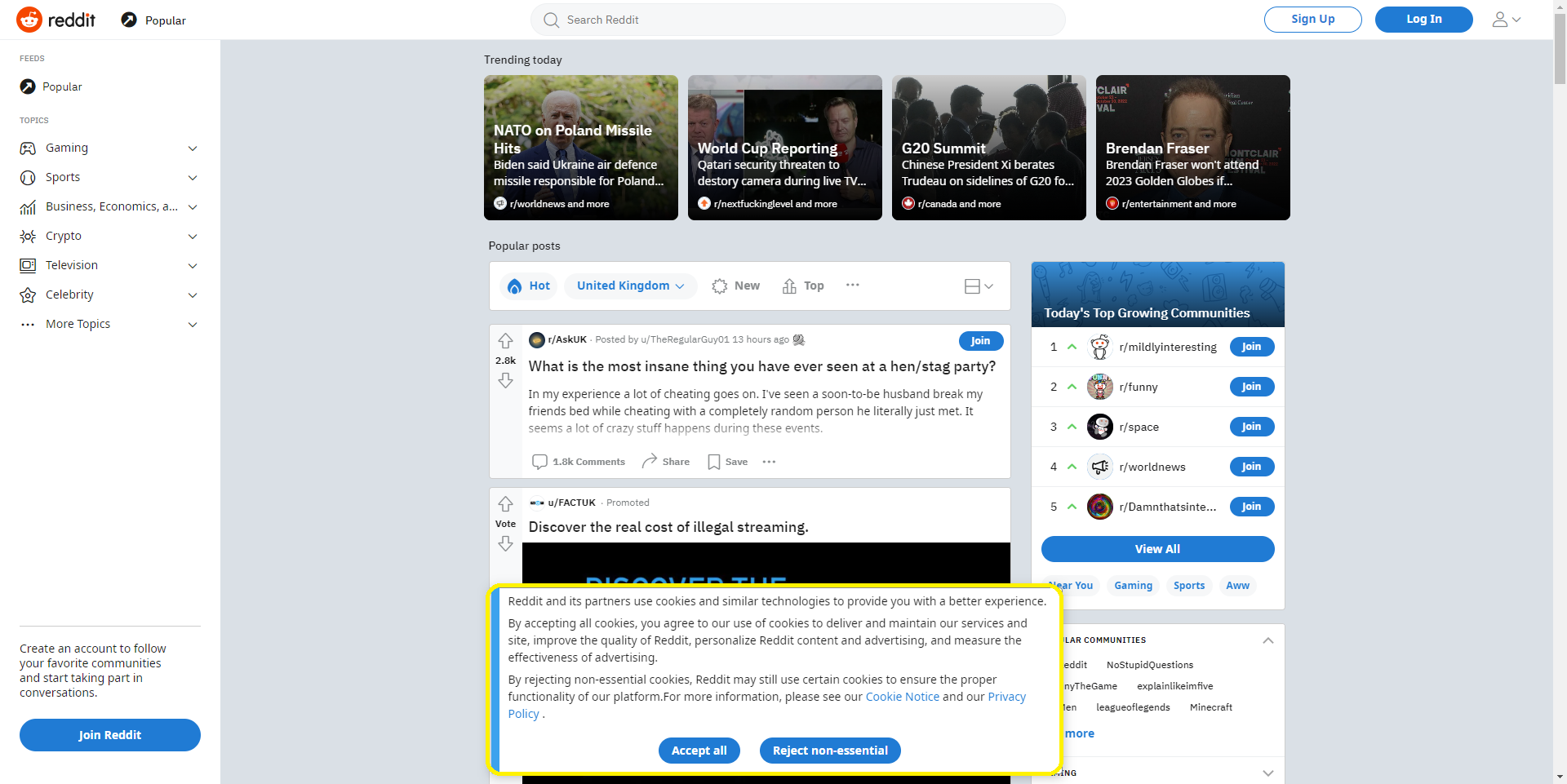}
    \alt{Visual representation of reddit.com, offering a cookie notice with two options: ``Accept all'' and ``Reject non-essential''. The cookie notice is at the bottom centre of the web page.}
    \caption{Visual representation of reddit.com cookie notice. None of the two screen readers could audibly output the cookie notice. They only read the body of the web page.}
    \label{fig:reddit}
\end{figure}

\begin{figure}[!htbp]
    \centering
    \scriptsize
    \begin{tabular}{|lp{7cm}|}
     \multicolumn{2}{c} {\includegraphics[scale=.47]{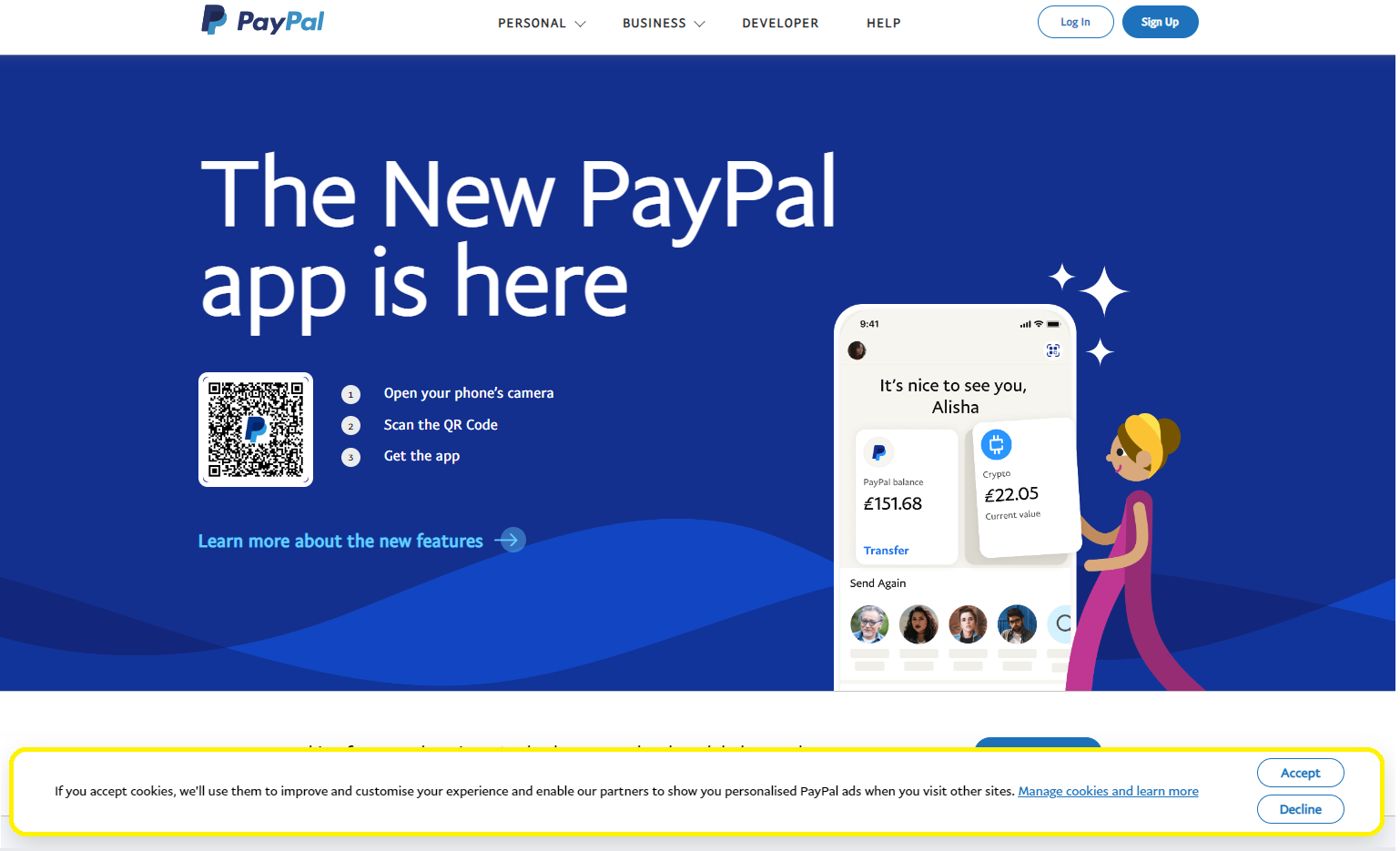}\alt{Visual representation of paypal.com, offering a cookie notice with two options: ``Accept'' and ``Decline''. The cookie notice is at the bottom of the web page.}}\\
        \multicolumn{2}{l}{(Enter webpage URL and press Enter Key)} \\\hline
        JAWS: & A simple and safer way to pay and get paid, vertical bar, PayPal UK \\
        JAWS: & Page has three regions, 8 headings and 33 links \\
        JAWS: & A simple and safer way to pay and get paid, vertical bar, PayPal UK \\
        JAWS: & Link PayPal \\
        JAWS: & Navigation region, list of four items \\
        \multicolumn{2}{|l|}{(Reads aloud navigation bar)}\\ 
        \multicolumn{2}{|l|}{(Navigate through body of site)} \\ 
        JAWS: & If you accept cookies, we'll use them to improve and customise your experience and enable our partners to show you personalised PayPal ads when you visit other sites. \\
        JAWS: & Link, manage cookies and learn more \\
        JAWS: & Accept button \\
        JAWS: & Decline button \\ \hline
    \end{tabular}
    \caption{Top: Graphical representation of paypal.com with highlighted cookie notice at the bottom of the page. Bottom: Example transcript via a screen reader.}\label{fig:paypal}
\end{figure}

\begin{figure}[!htbp]
    \centering
    \footnotesize
    \begin{tabular}{|lp{7cm}|}
       \multicolumn{2}{l} {\includegraphics[scale=.6]{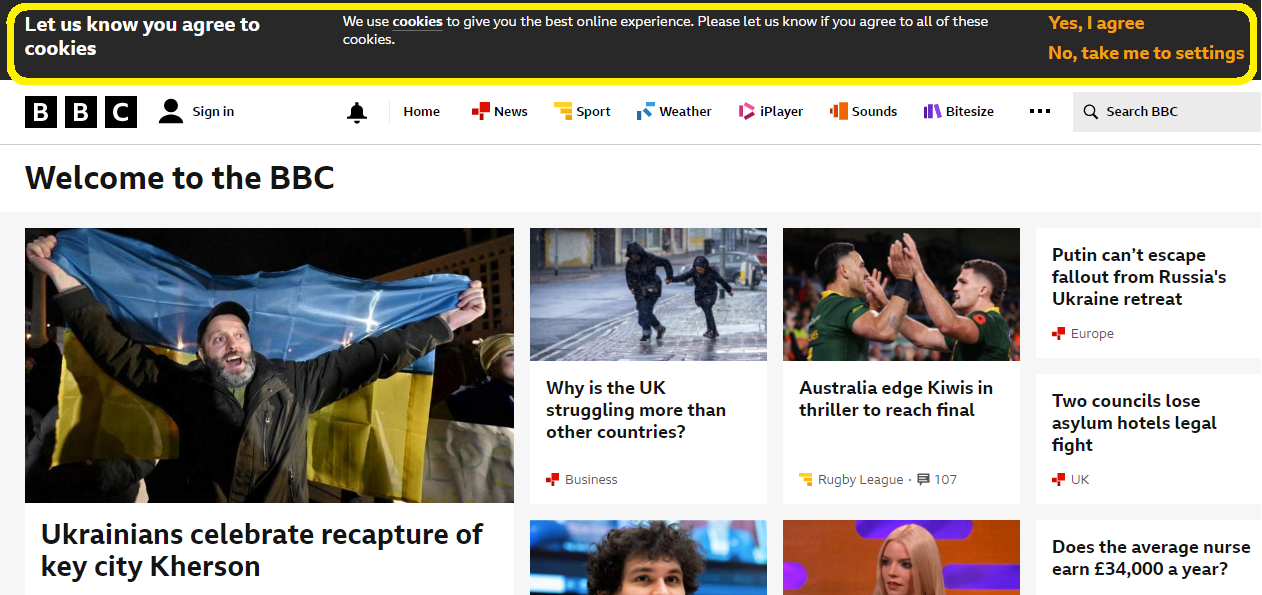}\alt{Visual cookie notice of bbc.co.uk, offering two options: ``Yes, I agree'' and ``No, take me to settings''. The cookie notice is located at the top of the web page.}}\\
        \multicolumn{2}{l}{(Enter webpage URL and press Enter Key)} \\\hline
        NVDA: & BBC.co.uk selected \\
        NVDA: & BBC Home \\
        NVDA: & Banner landmark, Let us know you agree to cookies, Heading level two \\
        NVDA: &  Clickable banner landmark, We use, link, cookies to give you the best online experience. Please let us know if you agree to all of these cookies \\
        NVDA: & Button, Yes, I agree \\
        NVDA: & Link, No, take me to settings \\
        NVDA: & BBC Navigation landmark, BBC Homepage \\\hline
        \multicolumn{2}{l}{(Proceeds to read the rest of the home page)}\\
    \end{tabular}
    \caption{Top: Graphical representation of bbc.co.uk with highlighted cookie notice at the top of the page. Bottom: Example transcript via a screen reader.}\label{fig:bbc}
\end{figure}
\newpage
\section{Answers to demographic questions}
\begin{table*}[h!]
    \centering
    \caption{Answers to demographic questions}
    \scriptsize
        \begin{tabular}{m{2.3cm}clclclc} \toprule
             \textbf{1.2 Devices used} & \textit{N} & \textbf{1.3 To Answer} & \textit{N} & \textbf{1.4 Hours a Week} & \textit{N}& \textbf{1.5 AT} & \textit{N} \\\midrule
            Personal Computer & 94 & Personal Computer & 71 & More than 30 hours & 28 & Magnification software & 50 \\
            Mobile Phone & 94 & Mobile Phone & 22 & 26-30 hours & 23 & Screen reader & 42 \\
            Tablet Computer & 47 & Tablet Computer & 5 & 6-10 hours & 14 & Assistive browser extension & 22\\
            Gaming consoles & 36 & Other & 2 & 16-20 hours & 12 & Other & 14\\
            Internet-enabled TV & 34 & & & 21-15 hours & 10 & None & 9 \\
            Smart home devices & 33 & & & 11-15 hours & 8 & Alternative input devices & 7 \\
            Wearable devices & 16 & & & 1-5 hours & 5 & Braille Display & 2 \\
            Public computers & 9 & & & & & Text Only Browser & 1  \\
            
            \bottomrule
            \textbf{1.6 Screen reader} & \textit{N} & \textbf{5.1 Age} & \textit{N}  & \textbf{5.2 Gender} & \textit{N} & \textbf{5.3 Education} & \textit{N} \\\midrule
            No screen reader & 40 &  25 to 34 & 33 & Male & 59 & Undergraduate  & 37 \\
            VoiceOver & 18 & 45 to 54 & 20 & Female & 38 & Graduate & 26 \\
            Narrator &  12 & 18 to 24 & 18 & Non-binary & 2 & Post-secondary  & 19 \\
            ChromeVox & 11 & 35 to 44 & 14 & Prefer not to say & 1 & Secondary  & 15 \\
            JAWS & 10 & 55 to 64 & 12 &&  & Prefer not to say & 3 \\
            Natural Reader & 10 & 65 or over & 2 \\
            Read\&Write & 9 & Prefer not to say & 1 \\
            Talkback & 6 & \\
            NVDA & 5 & \\
            Other & 5 & \\
            Orca & 1 & \\

            \bottomrule
            \textbf{5.5 Online Services} & \textit{N} & \textbf{1.7 Experience} & \textit{N}\\\hline
            Email & 98 & Basic & 33 \\
            Shopping and e-commerce & 93 & Moderate & 29 \\
            Social media & 90 &  None & 20 \\
            Payment/banking & 89 & Expert  & 9\\
            Video Streaming & 80 &  Previous experience & 5\\
            Music/audio streaming & 77 & Other  & 4 \\
            Travel booking & 68  \\
            File sharing and cloud storage & 64 \\
            Education and e-learning & 62 \\
            Commutation and collaboration & 35 \\
            \bottomrule
        \end{tabular}
        \label{tab:demographics}
    \end{table*}  
    
\end{document}